\documentclass[12pt]{article}
\pdfoutput=1

\usepackage[utf8]{inputenc}
\usepackage[left=2.55cm, right=2.55cm, top=2.55cm, bottom=2.55cm]{geometry}
\usepackage{amsmath,amssymb,amsbsy}
\usepackage{slashed}
\usepackage{xcolor}
\usepackage{graphicx}
\usepackage{url}
\usepackage{cancel}
\usepackage{cite}
\usepackage[colorlinks=true,allcolors=blue,pdfborder={0 0 0},linktocpage=false,pdfencoding=auto]{hyperref}
\usepackage{tabularx,booktabs}
\usepackage{multicol}
\usepackage{feynmp}
\usepackage{units}
\usepackage{xspace}
\usepackage[labelfont=bf]{caption}
\usepackage[section]{placeins}
\usepackage{subcaption}
\usepackage{soul}
\usepackage{bbold}
\usepackage{parskip}
\usepackage{float}
\usepackage{tabulary}
\usepackage{ulem}

\definecolor{darkred}{rgb}{0.6,0,0}
\definecolor{darkpurple}{rgb}{0.5,0,0.5}

\newcommand{\beqn}{\begin{eqnarray}}
\newcommand{\eeqn}{\end{eqnarray}}
\def\non{\nonumber\\}
\def\O{{\cal{O}}}
\def\E{{\cal{E}}}
\def \V{{\cal{V}}}
\def\A{{\cal{A}}}
\def\G{{\cal{G}}}

\begin{document}
\author{Amin Aboubrahim$^a$\footnote{\href{mailto:aabouibr@uni-muenster.de}{aabouibr@uni-muenster.de}}~~and Pran Nath$^b$\footnote{\href{mailto:p.nath@northeastern.edu}{p.nath@northeastern.edu}}
 \\~\\
$^{a}$\textit{\normalsize Institut f\"ur Theoretische Physik, Westf\"alische Wilhelms-Universit\"at M\"unster,} \\
\textit{\normalsize Wilhelm-Klemm-Stra{\ss}e 9, 48149 M\"unster, Germany} \\
$^{b}$\textit{\normalsize Department of Physics, Northeastern University,} \\
\textit{\normalsize 111 Forsyth Street, Boston, MA 02115-5000, USA} \\
}

\title{\vspace{-2cm}\begin{flushright}
{\small MS-TP-22-15}
\end{flushright}
\vspace{1cm}
\Large \bf
A tower of hidden sectors: a general treatment and physics implications
 \vspace{0.5cm}}
\date{}
\maketitle

\begin{abstract}

An analysis of a tower of hidden sectors coupled to each other, with one of these hidden sectors coupled to the visible sector, is given and the implications of such
couplings on physics in the visible sector are investigated. Thus the analysis considers  $n$  number of hidden sectors where the visible sector couples only to hidden sector 1, while the latter couples also to hidden sector 2, and the hidden sector 2 couples to hidden sector 3 and so on.  A set of successively feeble couplings of the hidden sectors to the visible sector are generated in such a set up. In general 
each of these sectors live in a different heat bath. We develop a closed form set of coupled Boltzmann equations for the correlated evolution of the temperatures and number densities of each of the heat baths. We then  apply the formalism to a simplified model with scalar portals between the different sectors. Predictions related to dark matter direct detection experiments and future CMB probes of dark radiation are made. 

\end{abstract}

\numberwithin{equation}{section}

\newpage

{  \hrule height 0.4mm \hypersetup{colorlinks=black,linktocpage=true} \tableofcontents
\vspace{0.5cm}
 \hrule height 0.4mm}

\section{Introduction}\label{sec:intro}

Most modern theories of particle physics based on supergravity, strings
and branes contain hidden sectors which are neutral under the Standard Model (SM) gauge
 group but may possess their own gauge groups. However, even though
 the hidden sectors are neutral under the SM gauge group, they may
  possess feeble interactions with the visible sector via a variety of portals
  such as the Higgs portal, kinetic and Stueckelberg mass mixings
  between the hidden sector $U(1)$ gauge fields and the visible sector $U(1)_Y$ hypercharge gauge field, or via higher dimensional operators
  which are dually charged under the gauge groups of the Standard Model 
  and of the hidden sectors.  The visible and the hidden sectors will in 
  general reside in different heat baths.
 In prior literature, it is often assumed (see, e.g.~\cite{Ackerman:2008kmp,Heurtier:2019git})
  that the visible and the hidden sectors temperature evolution are governed
  by entropy conservation carried out separately for the visible and the hidden sectors.
  However, such an assumption is untenable if there is a coupling between
  the hidden sectors and the visible sector. Thus a proper 
treatment of the  temperatures of the visible and the hidden sectors implies that  one perform their evolution consistent with conservation of the total
entropy and not the entropy conservation for each sector separately.
  An analysis of such a formalism was given in ref.~\cite{Aboubrahim:2020lnr}  
  for the case of 
  one hidden sector and for the case of two hidden sectors in ref.~\cite{Aboubrahim:2021ycj}.
  Further, an application of ref.~\cite{Aboubrahim:2020lnr}  
  was made for the explanation of
  EDGES anomaly~\cite{Bowman:2018yin} in ref.~\cite{Aboubrahim:2021ohe}
  (for related works see  \cite{Munoz:2018pzp,Munoz:2018jwq,Liu:2019knx,Berlin:2018sjs}).
     It is to be noted that the extension to two hidden sectors
  brings in new physics. Thus with one hidden sector, it is not possible
  to have dark photon as dark matter because it is not possible to satisfy
  the twin constraints that the dark photon simultaneously have a lifetime
  larger than the lifetime of the universe and at the same give us significant
  amount of dark matter consistent with the Planck 
  data~\cite{Planck:2018vyg,Aghanim:2018eyx}.
    We expect that further extensions of the formalism involving more hidden
    sectors would also bring in new physics which may have implications for particle physics and cosmology.
  
   Thus in this work we extend the analysis to the case of $n$ number
   of hidden sectors which are linked to the visible sector in a sequential
   manner. Specifically,  if we label the visible and the hidden sectors
   as $S_a, a=0,1,\cdots,n$ where $S_0$ stands for the visible sector
   and $S_a, a=1,\cdots, n$ stand for the hidden sectors, then we
   assume that the sectors have only next neighbor couplings. That is
   to say that  a sector $S_i$ has couplings only with $S_{i-1}$ and $S_{i+1}$.
    In this setup, we investigate the correlated evolution of temperatures
    in each sector relative to a given reference temperature.  While such a reference temperature can be chosen arbitrarily, in this analysis
    we assume the temperature of the visible sector to be the
    reference temperature or the `clock'.
   Specifically, we obtain a set of $n$ coupled Boltzmann equations for the temperature evolution functions $\xi_a= T_a/T, a=1,2,\cdots,n$,
   where $T$ is the temperature of the visible sector and $T_a$ ($a=1,\cdots,n$)
   is the temperature of hidden sector $a$.   
   An analysis is also given of how very feeble couplings can
   be obtained by a sequential couplings of the hidden sectors to the 
   visible sector. For some early works related to the 
   evolution of hidden  and visible sectors see, e.g.,~\cite{Foot:2014uba,Foot:2016wvj}.
  
 The outline of the rest of the paper is as follows:  In section~\ref{sec:model},
  we discuss the model with $n$ hidden sectors, and in section~\ref{sec:theoryapp} we 
  apply this generalization to $n$ hidden sectors to the Stueckelberg extension of the SM. In section~\ref{sec:numapp} we discuss the special case of three hidden sectors for a simple model consisting of a dark Dirac fermion and dark scalar particles. 
  In section~\ref{sec5} we  study the thermal effect of hidden sectors on experimentally measured observables such as the spin-independent proton-DM scattering cross-section and on dark radiation given by   
   $\Delta N_{\rm eff}$ for extra light relics. Conclusions are given in section~\ref{sec:conc}. A detailed derivation of the evolution 
   equations for visible and hidden sector temperatures is given in Appendix~\ref{app:A} and the collision and source terms used in the Boltzmann equations are listed in Appendix~\ref{app:B}.

\section{A model  with $n$ hidden sectors}\label{sec:model}

\begin{figure}[h]
\centering
\includegraphics[width=0.85\textwidth]{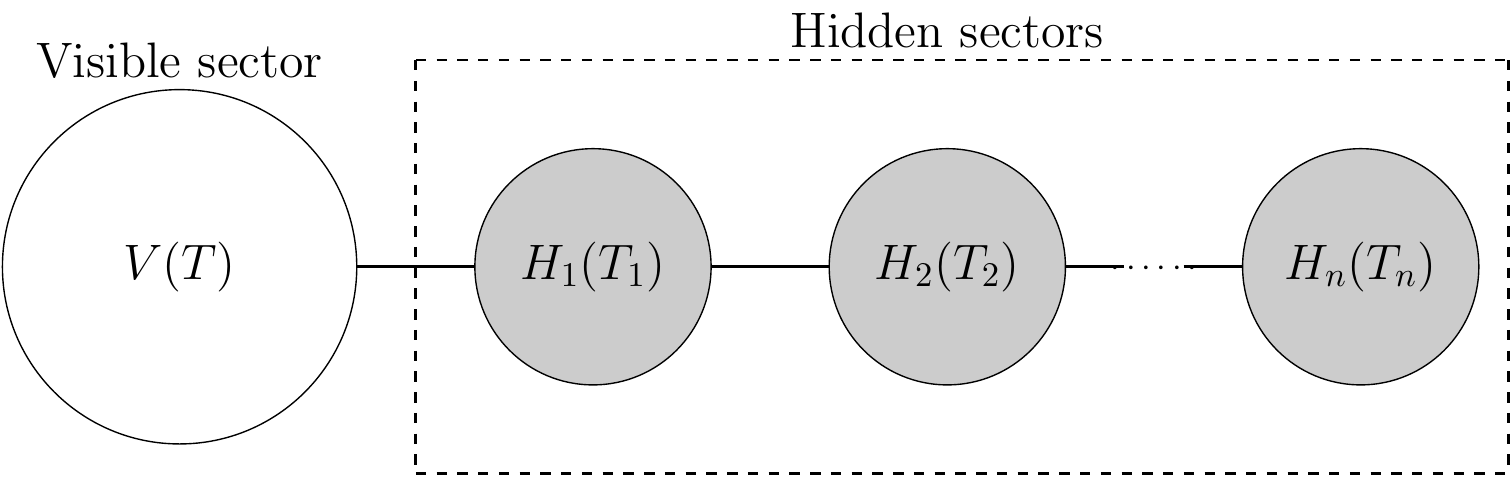}
\caption{The visible sector linked to a chain of $n$ number of hidden sectors. }
\label{figa}
\end{figure}

The model consists of  $n$ number of hidden sectors linked in a chain with the first hidden sector connected to 
the visible sector as shown in Fig.~\ref{figa}.

The above is a generalization of the cases considered previously for $n=1$ and $n=2$.
As mentioned earlier, we choose the temperature of the visible sector as the clock.
The set of Boltzmann equations for the energy density $\rho_\alpha$ including energy exchange between the sectors is $d\rho_\alpha/dt+3H(p_\alpha+\rho_\alpha)=j_\alpha$ ($\alpha=0,1,2,\cdots,n$). For the range of masses for the dark sector particles we consider in this analysis, it is safe to assume radiation domination\footnote{Early matter domination is possible, e.g., in the presence of heavy long-lived particles and rapidly oscillating fields (such as the inflaton field). In string theory, oscillating moduli fields displaced from the minimum of their potential can dominate the energy density in the early universe. None of these situations are relevant in our analysis. The light spectrum of our dark species justify the radiation domination assumption. } down to $T\sim 10^{-5}$ GeV and so $p_\alpha=1/3 \rho_\alpha$. Thus, we get
\begin{align}
\frac{d\rho_\alpha}{dt} + 4 H\rho_\alpha&=j_\alpha,~~\alpha=0,1,2,\dots,n\,,
\label{2.a} 
\end{align}
where $\alpha=0$ refers to the visible sector and
where $j_\alpha$ encodes in it all the possible processes exchanging energy between neighboring sectors. 
As discussed above, in order to properly describe the thermal evolution of the 
universe with visible and hidden sectors in different heat baths, we need to 
to have evolution equations for the $d\xi_\alpha/dT$. However, to 
 compute $d\xi_\alpha/dT, \alpha=1,2,\cdots, n$ we first derive 
 equations for $d\rho_\alpha/dT~ (\alpha=1\cdots n)$ in terms of 
$d\rho_v/dT$. We proceed by considering the Boltzmann equation of the total energy density $\rho$, i.e., $d\rho/dt+ 4H\rho=0$. Knowing that $d\rho/dt=(d\rho/dT)(dT/dt)$, we have
\begin{align}
 \frac{dT}{dt}&=- \frac{4H\rho}{d\rho/dT}.
 \label{dT-dt}
 \end{align}
 Next we look at the Boltzmann equations for the individual energy densities 
 $d\rho_\alpha/dt+ 4H\rho_\alpha=j_\alpha$ which after using Eq.~(\ref{dT-dt}) will immediately give us  
 \begin{align}
 \frac{d\rho_\alpha}{dT}=\frac{(4H\rho_\alpha-j_\alpha)}{4H\rho} \frac{d\rho}{dT}.
 \label{dr-dT}
 \end{align}
Next we define 
\begin{equation}
\begin{aligned}
\sigma_v&=\rho-\rho_v, \\
\sigma_\alpha&=\rho-\rho_\alpha=  \rho_v+ \sum_{\beta\neq \alpha} 
\rho_\beta \quad (\alpha,\beta=1\cdots n),
\end{aligned}    
\end{equation}
where the total energy density is
\begin{equation}
\rho= \rho_v+ \sum_{\alpha=1}^n \rho_\alpha\,.    
\end{equation}
To reduce clutter, we define 
$K_\alpha\equiv (4H\rho_\alpha-j_\alpha)/4H\rho$ and so Eq.~(\ref{dr-dT}) becomes
\begin{align}
\frac{d\rho_\alpha}{dT}&= K_\alpha \frac{d\rho}{dT} 
 = K_\alpha\left(\frac{d\rho_v}{dT}+ \frac{d\rho_\alpha}{dT}+ \sum_{\beta\neq \alpha} \frac{d\rho_\beta}{dT}\right). 
\end{align}
Rearranging, we get
\begin{align}
\frac{d\rho_\alpha}{dT} - C_\alpha \sum_{\beta\neq \alpha} \frac{d\rho_\beta}{dT}= 
C_\alpha \,\frac{d\rho_v}{dT},
\end{align}
where we have defined
\begin{equation}
C_\alpha\equiv \frac{ K_\alpha}{1-K_\alpha} = \frac{4H\rho_\alpha-j_\alpha}{4H\rho- 4H\rho_\alpha+j_\alpha} =\frac{4H\rho_\alpha-j_\alpha}{4H\sigma_\alpha+j_\alpha}.    
\end{equation}
Further, we may write out the coupled equation for $d\rho_\alpha/dT$ as follows
\begin{align}
&d\rho_1/dT- C_1 d\rho_2/dT -C_1 d\rho_3/dT \cdots -C_1 d\rho_n/dT =C_1 d\rho_v/dT, \non
&-C_2d\rho_1/dT + d\rho_2/dT -C_2 d\rho_3/dT \cdots -C_2 d\rho_n/dT =C_2 d\rho_v/dT, \non 
&\cdots\cdots\non
&-C_nd\rho_1/dT  -C_nd\rho_2/dT-C_n d\rho_3/dT \cdots  +  d\rho_n/dT =C_n d\rho_v/dT.
\end{align}
We can write this in a matrix form so that 
\begin{align}
\left(\begin{matrix} 1& -C_1&-C_1&\cdots& -C_1 & -C_1\\
-C_2& 1&-C_2&\cdots& -C_2 & -C_2\\
-C_3& -C_3&1&\cdots& -C_3 & -C_3\\
\cdots& \cdots &  \cdots& \cdots & \cdots& \cdots \\ 
-C_{n-1}& -C_{n-1}&-C_{n-1}&\cdots& 1 & -C_{n-1}\\
-C_n& -C_n&-C_n&\cdots& -C_n & 1\end{matrix} \right)
\left(\begin{matrix} d\rho_1/dT\\ d\rho_2/dT \\ d\rho_3/dT \\ \cdots \\ d\rho_{n-1}/dT \\ d\rho_n/dT \end{matrix} \right)=
\left(\begin{matrix} C_1\\ C_2\\ C_3\\ \cdots \\ C_{n-1} \\ C_n\end{matrix} \right) 
d\rho_v/dT.
\label{rho-eq1}
\end{align}
This equation can be written in a compact form so that
\begin{align}
\sum_{\beta=1}^n {\cal C}_{\alpha\beta} \frac{d\rho_\beta}{dT} = C_\alpha \frac{d\rho_v}{dT},
 \end{align}
where the matrix ${\cal C}$ is the square matrix on the left hand side of Eq.~(\ref{rho-eq1}).
 Thus $d\rho_\alpha/dT$ can be written as 
\begin{align}
\frac{d\rho_\alpha}{dT}=\sum_{j=1}^n ({\cal C}^{-1})_{\alpha\beta} C_\beta \frac{d\rho_v}{dT}.
\label{dridT}
\end{align}
Note that one may also write 
\begin{align}
\rho_\alpha'&=\frac{d\rho_\alpha}{dT}= P_\alpha + Q_\alpha \xi_\alpha', ~~~\alpha=1,2,\cdots, n-1, n,
\label{2.} 
\end{align}
where 
 $ \xi'_\alpha\equiv d\xi_\alpha/dT$ and where $P_\alpha$ and $Q_\alpha$ are 
 defined so that
 \begin{align}
  P_\alpha= \xi_\alpha \frac{dP_\alpha}{dT_\alpha}, ~~
Q_\alpha= T \frac{d\rho_\alpha}{dT_\alpha}.
\label{PQ}
\end{align}
Thus we have an equation for $d\xi_\alpha/dT$ which takes the form
\begin{align}
\frac{d\xi_\alpha}{dT}=-\frac{P_\alpha}{Q_\alpha} +     \sum_{\beta=1}^n(C^{-1})_{\alpha\beta} C_\beta\frac{\rho_v'}{Q_\alpha},~~~~\alpha=1,2,\cdots, n-1,n.
\label{2.a10} 
\end{align}
Eqs.~(\ref{2.a10}) give us $n$ differential equations for the evolution functions $d\xi_\alpha/dT$. These have to be solved along with the Boltzmann equations governing the number density evolution of the hidden sector particles. This will allow us to determine the relic densities of all stable species and describe the thermal evolution of this coupled system.

For the special case of three hidden sectors, the matrix ${\cal{C}}$ is given by 
 \begin{align}
{\cal C}=\left(\begin{matrix} 1& -C_1&-C_1\\
-C_2& 1&-C_2\\
-C_3& -C_3&1\end{matrix} \right).
 \end{align}
Therefore we can write Eq.~(\ref{dridT}) in a matrix form so that 
\begin{align}
\left(\begin{matrix} d\rho_1/dT\\ d\rho_2/dT \\ d\rho_3/dT \end{matrix} \right)=
\frac{1}{D} \left(\begin{matrix} (1-C_2C_3)& C_1(1+C_3)& C_1(1+C_2)\\
C_2(1+C_3) & (1-C_1C_3) & C_2(1+C_1) \\
  C_3(1+C_2)& C_3(1+C_1) &(1-C_1C_2)
  \end{matrix} \right)
\left(\begin{matrix} C_1\\ C_2\\ C_3\end{matrix} \right) 
d\rho_v/dT,
\end{align}
where $D$ is given by 
 \begin{align}
 D&= 1-C_1C_2-C_1C_3- C_2C_3 -2 C_1 C_2 C_3.
 \label{det}
 \end{align} 
Since $\rho_\alpha'= P_\alpha+ Q_\alpha \xi_\alpha'$, we have
\begin{align}
\xi_\alpha'\equiv\frac{d\xi_\alpha}{dT}= - \frac{P_\alpha}{Q_\alpha} + \frac{1}{Q_\alpha} \rho'_\alpha.
\end{align}
Therefore the individual expressions can be easily obtained as
\begin{align}
\frac{d\xi_1}{dT}&=- \frac{P_1}{Q_1}+ 
 \frac{1}{Q_1D} \left[C_1+ C_1C_2+ C_1C_3 + C_1C_2C_3)\right] 
 d\rho_v/dT, \\
\frac{d\xi_2}{dT}&=- \frac{P_2}{Q_2}+ 
 \frac{1}{Q_2D} \left[C_2+ C_1C_2+ C_2C_3 + C_1C_2C_3)\right] 
 d\rho_v/dT, \\
\frac{d\xi_3}{dT}&=- \frac{P_3}{Q_3}+ 
 \frac{1}{Q_3D} \left[C_3+ C_2C_3+ C_1C_3 + C_1C_2C_3)\right] 
 d\rho_v/dT.
 \end{align}
 These equations give the  evolution of the temperature ratios
$\xi_\alpha=T_\alpha/T, \alpha=1,2,3$. 

Before proceeding further, we note here that a chain of $n$ hidden sectors analogous to the one of Fig.~\ref{figa} appears in well known scenarios of physics beyond the Standard Model such  as in moose/quiver gauge theories (see, e.g.,~\cite{Rothstein:2001tu,Douglas:1996sw,Arkani-Hamed:2001kyx,Hill:2000mu}). 
 Thus an analysis of the type above may find application in a broader class of  
  particle physics
 models beyond the Standard Model.

\section{Theory application: The Stueckelberg extension of the Standard Model}\label{sec:theoryapp}

The Stueckelberg extension of the SM with an extra hidden $U(1)_X$ gauge group offers the possibility of a hidden sector feebly coupled to the visible sector. The gauged $U(1)_X$ gives rise to a dark photon which, under certain conditions, can be a dark matter candidate~\cite{Aboubrahim:2021dei,Aboubrahim:2021ycj}. In this section we give the analytic expressions for a Stueckelberg extension of the SM to $n$ hidden sectors which the above formalism can easily be applied to. 

\subsection{Generation of very feeble couplings from kinetic mixings via a chain of hidden sectors}

As discussed in section~\ref{sec:model}, we
 consider the following type of couplings: the visible sector couples to the  hidden sector 1 $(H_1)$,
the hidden sector 1 couples to hidden sector 2 $(H_2)$ 
 and so on, up to hidden sector ($n-1$) which
couples to the hidden sector $n$ $(H_n)$.
The couplings between the sectors could be via kinetic mixing~\cite{Holdom:1985ag},  via Stueckelberg mass mixing~\cite{Kors:2004dx}, via both kinetic mixing and Stueckelberg 
mass mixing~\cite{Feldman:2007wj} and contain matter in the  hidden sectors~\cite{Kors:2004dx,Cheung:2007ut} 
(for recent works based
on Stueckelberg extensions of the Standard Model see also~\cite{Aboubrahim:2019kpb,Du:2022fqv}).
Specifically we assume that each of the $n$ number of hidden sectors possess a $U(1)_{X_I}$ gauge 
invariance with a gauge field $A_I^\mu, I=1,\cdots,n$. Including the gauge field $B_\mu$ for the
hypercharge $U(1)_Y$, we have $n+1$ $U(1)$ gauge fields which we collectively denote
by $V_i^\mu, i=1,\cdots,n+1$ where we define the elements of $V_i$ as follows: 
 \begin{align}
   (V)=(V_1=B, V_{2}= A_{1},\cdots,V_{n+1}=A_{n}),
  \end{align}
  where $B_\mu$ is the hypercharge field of the visible sector and 
  $A_{I\mu}, ~I=1,\cdots,n$ are the  gauge fields for  the 
  hidden sector   $U(1)_{X_I}, ~I=1,\cdots,n$.
  We assume the gauge fields have kinetic mixings and Stueckelberg mass
  mixings
  and  such mixings to be sequential
  and thus the Lagrangian for the gauge fields 
  is assumed to be of the form 
 \begin{align}
 \mathcal{L}_{\rm km}=& -\frac{1}{4}\left[  
 \sum_{i=1}^{n+1} G_{i\mu\nu}G_i^{\mu\nu} 
 +2\sum_{i=1}^{n} \delta_{i} G_{i\mu\nu} G_{i+1}^{\mu\nu} \right]\non
& -\frac{1}{2} 
   \sum_{i=1}^{n} 
   (M_{i} V_{i\mu} + \mu_i V_{i+1\mu} +\partial_\mu\sigma_{i})^2.
\label{2.} 
\end{align}
Note that there are $n$ axion fields $\sigma_i, ~i=1,\cdots, n$ which are absorbed by the
$n$ gauge fields of the hidden sectors to become massive.

We note that the fields in the Lagrangian above are  not in their canonical
form. Further, 
this Lagrangian must be coupled with the Standard Model Higgs field
in the electroweak symmetry breaking. 
After the electroweak symmetry breaking, one must make the
appropriate transformations to put the Lagrangian in the canonical form. We give now a brief discussion of this procedure.
 In the analysis below we will include the third component of the gauge fields for $SU(2)_L({\cal{A}}_3)$ 
of the SM
since that fields along with the hypercharge gauge field enter in 
giving mass to the $W$ and $Z$ bosons in the electroweak symmetry breaking.
 The transformation that puts the kinetic energy in the 
canonical form is given by 
\begin{align}
\V_\mu = K \V'_\mu\,,
\label{kdefined}
\end{align}
where our notation is the following\footnote{Note that while $V_{i\mu}$ has $i$ taking values 
$i=1,\dots,n+1$, $\V_a$ has $a$ taking values $a=0,1,\cdots, n+1$.}
\begin{equation}
\begin{aligned}
\V_\mu^T&=(\{\V_a\})=
(\A_{3\mu}, V_{1\mu}, V_{2\mu}, V_{3\mu},\cdots,  V_{n\mu}, V_{n+1,\mu}), 
a=0,\cdots,n+1, \non
\V_\mu^{'T} &=(\{\V'_a\})=
(\A_{3\mu}, V'_{1\mu}, V'_{2\mu}, V'_{3\mu},\cdots, V'_{n\mu},
V'_{n+1,\mu}),a=0,\cdots,n+1,
\end{aligned}
\end{equation}
and where $K$ is an $(n+2)\times (n+2)$ dimensional matrix defined as
 \begin{align}
  K & = \left(\begin{matrix} 
 1 & 0 & 0& 0&0&\cdot&\cdot&0 \\ 
   0& 1 &-s_1 & s_1s_2& -s_1s_2 s_3 & \cdot  & \cdot & (-1)^n s_1s_2\cdots s_n
     \\ 
0&    0 &c_1& -c_1s_2& c_1s_2 s_3 & \cdot &\cdot & (-1)^{(n-1)} c_1 s_2 \cdots s_n\\
0&    0& 0& c_2& -c_2s_3 & \cdot & \cdot & (-1)^{(n-2)} c_2 s_3 \cdots s_n\\
 \cdot&   \cdot & \cdot &\cdot &\cdot & \cdot &\cdot& \cdot\\
  0&  0 & 0 &0 &0 & 0&  c_{(n-1)}   & -c_{(n-1)} s_n\\    
  0&     0 & 0 &0 &0 & 0 &0& c_n              
      \end{matrix}\right),
       \label{diag-10}
          \end{align}
where in general $c_k$ and $s_k$ (for $k=1,\dots, n$)  are given by
\begin{align}
    c_k=&
     \frac{\sqrt{1-\sum_{i=1}^{k-1}} \delta_i^2}{\sqrt{1-
     \sum_{i=1}^k}\delta_i^2},~~s_k=
     \frac{\delta_k}{\sqrt{1-
     \sum_{i=1}^k}\delta_i^2}.    
       \label{diag-7}     
       \end{align}
Note that $c_k, s_k$ satisfy the relation $c_k^2-s_k^2=1$. In the transformed
coordinates the kinetic energy part of the Lagrangian takes the form 
$\mathcal{L}= -\frac{1}{4} \V^{'T}_{\mu\nu} \V^{'\mu\nu}.$ 
Next, we focus on the mass terms. In addition to the Stueckelberg mass terms involving the visible
and the hidden sectors, we have mass terms for $A_0, {{\cal{A}}_3}$
from the Higgs mechanism which have the form 
\begin{align}
\mathcal{L}^m_{\text{SM}} = -\frac{1}{2}\left(\frac{1}{4} g_Y^2 v^2 B^\mu B_\mu
+    \frac{1}{4} g_2^2 v^2 {\cal{A}}^\mu_3\A_{3\mu} +  \frac{1}{2} g_1g_2 v^2B^\mu
{\cal{A}}^\mu_3\right),
\end{align}
 where $v= (\sqrt 2 G_F)^{-1/2}$ and where $G_F$ is the Fermi constant.
 Thus we write the total mass  term for the gauge fields  so that
 \begin{align}
 \mathcal{L}_m= - \frac{1}{2} \V_\mu^{T} M^2 \V^\mu,
 \end{align}
 where $M^2$ includes the mass terms both from the Higgs mechanism
 and from the Stueckelberg mass growth mechanism. After the transformation
 that diagonalizes the kinetic energy, the mass term is given by 
  \begin{align}
 \mathcal{L}_m= - \frac{1}{2} \V_\mu^{\prime T} M^{\prime 2} \V^{\prime\mu},
 \end{align} 
 where $\V^{'\mu}$ is defined in Eq.~(\ref{kdefined}) and $M^{\prime 2}= K^T M^2 K$.
  Since $M^{\prime 2}$ is a symmetric matrix it can be diagonalized by an
  orthogonal transformation $R$ so that $R^TR=I$.  Thus the transformation
  $\V^\prime= R\E$, or $\V= KR \E$ diagonalizes the mass matrix. 
    \begin{align}
  M_D^2&= {\mathcal{O}}^T M^2 {\mathcal{O}},\non
      \V&= {\mathcal{O}}\E, 
  \end{align}  
  where $\O= KR$ and 
  both the kinetic and the mass terms are in a canonically diagonal basis and we
  have 
\begin{align} 
  \mathcal{L}= - \frac{1}{4} {\E}^T_{\mu\nu} {\E}^{\mu\nu} - \frac{1}{2}
  {\E}^T_{\mu} M_D^2 \E^{\mu}. 
  \label{canonical}
 \end{align} 
  We note that the subcase $n=2$ was 
  discussed in~\cite{Aboubrahim:2020afx}. We arrange the vector  mass eigenstates $\E_\mu$
  so that $\E^T=(Z, A_{\gamma}, A_{\gamma_2}, \dots, A_{\gamma_3}, \cdots, A_{\gamma_{n}}$). 
  Thus the mass eigenstates consist of the $Z$ boson, the photon, and $n$ number of dark photons
  in the hidden sector.

\subsection{Neutral currents of the coupled visible and hidden sectors} 

Since the mixings involve the neutral gauge bosons of the 
      visible and the hidden sectors, we begin by displaying the
      interactions of these gauge bosons in their respective sectors.
      Thus the neutral current interactions of the visible and the hidden sectors together are 
\begin{align}
  \mathcal{L}_{\rm NC}= g_2 J_2^\mu {\cal{A}}_{3\mu}+ g_Y J_Y^\mu B_\mu
       + \sum_{i=1}^n g_{Xi} J^\mu_{Xi} A_{i\mu}.  
      \end{align} 
    In the equation above, matter charged under a hidden sector gauge
      field $A_{i\mu}$ will reside in the source term $J^\mu_{Xi}$. Such matter may
      consist of dark fermions and dark complex scalar fields charged under the
      $U(1)_{Xi}$. Thus if we have Dirac fermions $D_{Xi}$ charged under 
      $U(1)_{Xi}$ we will have $J^{\mu}_{Xi}= \bar D_{Xi}\gamma^\mu D_{Xi}$.       
      
To write the neutral currents in a compact notation in the $\V$ basis, we define
\begin{align}
      (\V_a):& ({\cal{A}}_{3\mu}, B_\mu, A_{1\mu},\cdots, A_{n\mu}),\non   
 (\G_a):&      \G_0=g_2, ~\G_1= g_Y, ~\G_{3}= g_{X_1},\cdots, \G_{n+1}=g_{X_n} ~a=0,\cdots,n+1,\non
  ({\cal{J}}_a): &   {\cal {J}}^\mu_0=J^\mu_2 ,
      {\cal {J}}^\mu_1= J^\mu_Y,\cdots,  {\cal {J}}^\mu_{i+1}= J^\mu_{X_i}, ~a=0,\cdots, n+1 \non
 (\E_a):&  {\cal{E}}^\mu_0= Z^\mu,  ~{\cal{E}}^\mu_1 =A^\mu_\gamma,
 ~{\cal{E}}^\mu_2 =A^\mu_{\gamma_2},\cdots, 
 , ~{\cal{E}}^\mu_{n+1}=  A^\mu_{\tilde \gamma_n}, a=0,\cdots, n+1,   
\end{align}
     where $Z^\mu$ is the $Z$-boson field, $A^\mu_\gamma$ is the photon field,
     and $A^\mu_{\tilde \gamma_i},\cdots, A^\mu_{\tilde \gamma_n}$     
      are the fields for the dark photons. 
     We now rewrite the neutral current interaction so that 
      \begin{align}
        \mathcal{L}_{\rm NC} 
       =\sum_{a=0}^{n+1}  \G_a {\cal{J}}^\mu_a \V_a. 
      \end{align} 
     Next we write $\mathcal{L}_{\rm NC}$   
     in terms of the mass eigenstates $\E$ so that
      \begin{align}
       \mathcal{L}_{\rm NC}    
       =\sum_{a=0}^{n+1}  \G_a {\cal{J}}^\mu_a {\cal{O}}_{ab}
       {\cal{E}}_{b\mu}.  
      \end{align}         
     We can now display  the coupling of the $Z$ boson and the photon
     and for the dark photon in an explicit manner.        
   Thus we can decompose the couplings into several parts so that 
     \begin{align}
      \mathcal{L}_{\rm NC}
        = \mathcal{L}^{\rm NC}_{\rm SM} + \mathcal{L}_{Z\gamma D} + \mathcal{L}_{\text{SM}\tilde\gamma} + \mathcal{L}_{D\tilde\gamma},  
     \end{align}
     where 
     \begin{equation}
     \begin{aligned}
     \mathcal{L}^{\rm NC}_{\rm SM}&= \left(g_2 J_2^\mu  {\cal{O}}_{00} + g_Y J^\mu_Y {\cal{O}}_{10}\right)Z_\mu    +  \left(g_2 J_2^\mu  {\cal{O}}_{01} + g_Y J^\mu_Y {\cal{O}}_{11}\right)A^\gamma_\mu, \\
  \mathcal{L}_{Z{\gamma}D}&=\sum_{a=2}^{n+1} g_{Xa} J^\mu_{Xa} \mathcal{O}_{a0} Z_\mu            
      + \sum_{a=2}^{n+1} g_{Xa} J^\mu_{Xa} \mathcal{O}_{a1} A^\gamma_\mu,   \\
     \mathcal{L}_{\text{SM}\tilde \gamma} & =\sum_{i=1}^n \left(g_2 J_2^\mu  {\cal{O}}_{0(i+1)} + g_Y J_Y {\cal{O}}_{1(i+1)}\right)A^{\tilde\gamma_{i}}_{\mu},  \\
    \mathcal{L}_{D\tilde \gamma}&=\sum_{i,j=1}^{n} g_{X_{i}} J_{X_{i}} \mathcal{O}_{(j+1)(i+1)} A^{\tilde\gamma_{j}}_\mu.    
    \label{pieces}      
          \end{aligned}
     \end{equation}
     Here $\mathcal{L}^{\rm NC}_{\rm SM}$ is the modified neutral current of the Standard Model,
     $\mathcal{L}_{Z{\gamma}D}$ is the coupling of the SM gauge bosons
     $Z, A_\gamma$ with the dark sector, $\mathcal{L}_{\text{SM}\tilde \gamma}$ is the
     coupling of the dark photons with the SM sector and 
     $\mathcal{L}_{D\tilde \gamma}$ is the coupling of the dark photons with 
     the dark sources.   
    The various parts of the couplings of Eqs.~(\ref{pieces}) have the 
        following meaning. As stated $\mathcal{L}_{\rm SM}$ is the modified neutral current
        of the Standard Model. $\mathcal{L}_{Z{\gamma}D}$ and $\mathcal{L}_{\text{SM}\tilde{\gamma}}$
        enter in the freeze-in production of the hidden sector particles,
        $\mathcal{L}_{\text{SM}\tilde \gamma}$ enters in the decay of the dark photons,
        and $\mathcal{L}_{D\tilde \gamma}$ controls freeze-out in the dark sector,
        while $\mathcal{L}_{Z{\gamma}D}$ enters in the analysis of kinetic equilibrium 
        of the dark sources after chemical decoupling. 

\section{Numerical application: The visible sector coupled to three hidden sectors}\label{sec:numapp} 

\subsection{A simple model}

In this section we will use the general expression of Eq.~(\ref{2.a10}) to determine the temperature evolution of a chain of hidden sectors coupled to the visible sector as described by Fig.~\ref{figa}. The hidden sectors will eventually get populated by dark particle species and so the evolution of the temperature is coupled with the evolution of the particle number densities. In refs.~\cite{Aboubrahim:2020lnr,Aboubrahim:2021ycj} the cases of one and two hidden sectors have been considered in the context of a Stueckelberg $U(1)$ extension of the SM. Here, we add a third hidden sector but consider a simpler model to illustrate the effect of three hidden sectors on dark matter phenomenology and on cosmology.
Further, we assume that particles in the different sectors communicate via a scalar portal which will become evident next. 

Our set up of the matter content of the three hidden sectors is as follows:
The first hidden sector has a Dirac fermion $D$ (which serves as one of the possible dark matter candidates) and a pseudo-scalar $\phi_1$ (mediator). 
The field $\phi_2$ is a scalar while $\phi_3$ is a pseudo-scalar and are not 
particularly associated with any specific hidden sector. The model presented below
is phenomenological and  its UV completion remains to be explored.
The Lagrangian of the model we use for the rest of the analysis is
\begin{align}
\mathcal{L}&=\mathcal{L}_{\rm SM}+\Delta\mathcal{L},
\end{align}
where $\mathcal{L}_{\rm SM}$ is the SM Lagrangian and $\Delta\mathcal{L}$
is the Lagrangian for the hidden sector including portal interactions between the 
visible and the hidden sectors.
We note here in passing that we are considering the hidden sector to have no gauged $U(1)$ symmetry and for that reason there is no kinetic energy for the hidden sector gauge field or its interactions  included in $\Delta\mathcal{L}$
below. Thus $\Delta\mathcal{L}$  is given by
\begin{align}
\Delta\mathcal{L}&=\frac{1}{2}\sum_{i=1}^3(\partial_\mu \phi_i \partial^\mu \phi_i-m_i^2\phi^2_i)+\bar{D}(i\gamma^\mu\partial_\mu-m_D)D\nonumber\\
&~~~- y_{q}\phi_1 \bar{q}\gamma^5 q- y_{\ell}\phi_1 \bar{\ell}\gamma^5 \ell-y_D\phi_1\bar{D}\gamma^5 D 
 \nonumber \\
&~~~-\frac{\kappa_1}{3}\phi_1^2\phi_2-\frac{\kappa_2}{3}\phi_2\phi_3^2-\frac{1}{4}\sum_{i=1}^3 \lambda_i \phi_i^4-\frac{1}{4}\sum_i\sum_{j>i}\lambda_{ij}\phi_i^2\phi_j^2\,,
\end{align}
where the dark fermion $D$ resides in the hidden sector 1 and couples only to the field $\phi_1$ in the hidden sector 1.
The fields $\phi_1$ and $\phi_3$ as pseudoscalars and $\phi_2$ as CP-even scalar are  
chosen purely on phenomenological grounds which we use to illustrate the main features of the visible sector coupling to three hidden sectors.
For simplicity, we assume that $\phi_1$ only couples to the first generation of quarks and leptons via the Yukawa couplings $y_q$ and $y_{\ell}$ and to the $D$ fermions via $y_D$. Thus in this model, the field $\phi_1$ is the portal connecting the visible and 
hidden sectors. Further, for simplicity, we will assume  equality of the portal field coupling to the Standard Model quarks and leptons, i.e., $y_{q}=y_{\ell}=y_f$.
The size of the Yukawas could be very small. Thus, for example,
 a higher dimensional operator such as $\frac{H^\dagger H}{\Lambda^2} \bar q q$
 after electroweak symmetry breaking would give $y_q\sim 10^{-9}$ for $\Lambda=
 {\cal O}(10^6)$ GeV. A  string scale such as this can arise in Type IIB string theory.
The quartic couplings $\lambda_i$ do not play a role in the temperature and number density evolution while we assume the couplings $\lambda_{ij}$ are small compared to $\kappa_1$ and $\kappa_2$. 

\subsection{Evolution equations}  

In the set up described above, only the first hidden sector possesses direct couplings to SM (the visible sector) through the Yukawa couplings
$y_q$ and $y_{\ell}$ while the second  hidden sector connects to the first and 
to the third hidden sectors.
We assume that the visible sector is held at a temperature $T$ and the three hidden sectors at temperatures $T_1$, $T_2$ and $T_3$
 where one prescribes the initial values of the temperatures $T_i\ll T$ ($i=1,2,3)$. The Boltzmann equations for the number and total energy densities are given by
\begin{align}
\label{dn}
\frac{dn_r}{dt}+3Hn_r=\mathcal{Q}_r\,, \\
\frac{d\rho}{dt}+3H(\rho+p)=0,
\label{dr}
\end{align}
with $r\in\{D,\phi_1,\phi_2,\phi_3\}$. In the above, $\rho$ is the total energy density, $p$ is the total pressure, $H$ is the Hubble parameter and 
$\mathcal{Q}_r$ are the collision terms which include all possible particle number-changing processes. They are listed in Appendix~\ref{app:B}.

After defining the co-moving number density or \textit{yield}, $Y=n/\mathbb{s}$,  where $\mathbb{s}$ is the total entropy density and using Eqs.~(\ref{dn}) and~(\ref{dr}), we write
\begin{align}
\frac{dY_r}{dT}=&-\mathbb{s}\frac{d\rho/dT}{4H\rho}\mathcal{Q}_r,
\label{boltz}
\end{align} 
where $H$ is related to the energy density by the Friedmann equation
\begin{equation}
H^2=\frac{8\pi G_N}{3}\rho.
\end{equation}
One can define the individual energy and entropy densities in terms of effective number of degrees of freedom so that
\begin{align}
\rho&=\frac{\pi^2}{30}\left(g_{\rm eff}^v T^4+g_{1\rm eff} T_1^4+g_{2\rm eff} T_2^4+ g_{3\rm eff} T_3^4
\right), \\
\mathbb{s}&=\frac{2\pi^2}{45}\left(h_{\rm eff}^v T^3+h_{1\rm eff} T_1^3+h_{2\rm eff} T_2^3 +h_{3\rm eff} T_3^3\right).
\label{rho-s}
\end{align}
We note that $g_{\rm eff}$ and $h_{\rm eff}$ of the hidden sectors are defined as integrals which depend on the mass 
of the relevant particle and temperature.  In  appendix~\ref{app:A}, we exhibit the $m/T$ 
 dependence of the integrals.
In the above, we need to compute $T_i$ so that $T_i=\xi_i T$. 
Thus we need to  write out the 
evolution equations for $\xi_i$  in an explicit form.  
In this case we have three evolution functions $\xi_1=T_1/T,
\xi_2=T_2/T$ and $\xi_3=T_3/T$ and using Eq.~(\ref{2.a10}) we find that they obey the following equations 
\begin{align}
\label{xi1p} 
\frac{d\xi_1}{dT}&=- \frac{P_1}{Q_1}+ 
 \frac{1}{Q_1D} \left[C_1+ C_1C_2+ C_1C_3 + C_1C_2C_3)\right] 
 d\rho_v/dT, \\
\label{xi2p}
\frac{d\xi_2}{dT}&=- \frac{P_2}{Q_2}+ 
 \frac{1}{Q_2D} \left[C_2+ C_1C_2+ C_2C_3 + C_1C_2C_3)\right] 
 d\rho_v/dT, \\
\frac{d\xi_3}{dT}&=- \frac{P_3}{Q_3}+ 
 \frac{1}{Q_3D} \left[C_3+ C_2C_3+ C_1C_3 + C_1C_2C_3)\right] 
 d\rho_v/dT,
 \label{xi3p}  
 \end{align}
 where $D$ is given by
 \begin{align}
 D&= 1-C_1C_2-C_1C_3- C_2C_3 -2 C_1 C_2 C_3.
 \label{det}
 \end{align}
For the case of one hidden sector ($n=1$), we have $C_2=C_3=0$, $D=1$ and  $\xi_2$
and $\xi_3$ are absent. In this case Eq.~(\ref{xi1p}) reads 
\begin{align}
\frac{d\xi_1}{dT}=  -\frac{P_1}{Q_1} + C_0\frac{\rho_v'}{Q_1},
\label{2.47} 
\end{align}
which agrees with the result of ref.~\cite{Aboubrahim:2020lnr}. Next we look at the $n=2$ case (two hidden sectors). Here 
we set $C_3=0$ and $D=1-C_1C_2$, so that
\begin{align}
\frac{d\xi_1}{dT}&= -\frac{P_1}{Q_1} +  \frac {C_1+ C_2C_1}{1-C_1C_2}
 \frac{\rho_v'}{Q_1}.
\label{2.45} 
\end{align}
Similarly we have 
\begin{align}
\frac{d\xi_2}{dT}&= -\frac{P_2}{Q_2} + 
 \frac {C_1C_2 + C_2} {1-C_1C_2}\frac{\rho_v'}{Q_2}.
\label{2.45} 
\end{align}
These agree with the analysis of refs.~\cite{Aboubrahim:2021ycj,Aboubrahim:2021dei}.
We note here that $C_1, C_2, C_3$ are given by
\begin{align}
C_1&=
 \frac{4H\rho_1-j_1}{4H\sigma_{1}+j_v},~~
C_2=
 \frac{4H\rho_2-j_2}{4H\sigma_{2}+j_2},~~ 
C_3=
 \frac{4H\rho_3-j_3}{4H\sigma_{3}+j_3},
\label{2.ax} 
\end{align} 
where the source terms admit the relation $j_v+ j_1+j_2+j_3=0$. 
The source terms $j_1$, $j_2$ and $j_3$ are listed in Appendix~\ref{app:B}.
 Carrying out some further simplifications to Eqs.~(\ref{xi1p})$-$(\ref{xi3p}), we can write them as
\begin{align}
\label{xi1}
\frac{d\xi_1}{dT}&= -\frac{\xi_1}{T}+\left(\frac{4H\rho_1-j_1}{4H\rho_v+j_1+j_2+j_3}\right)\frac{d\rho_v/dT}{T\frac{d\rho_1}{dT_1}}, \\
\label{xi2}
\frac{d\xi_2}{dT}&= -\frac{\xi_2}{T}+\left(\frac{4H\rho_2-j_2}{4H\rho_v+j_1+j_2+j_3}\right)\frac{d\rho_v/dT}{T\frac{d\rho_2}{dT_2}}, \\
\label{xi3}
\frac{d\xi_3}{dT}&= -\frac{\xi_3}{T}+\left(\frac{4H\rho_3-j_3}{4H\rho_v+j_1+j_2+j_3}\right)\frac{d\rho_v/dT}{T\frac{d\rho_3}{dT_3}}.
\end{align}
The factor $\frac{d\rho/dT}{4H\rho}$ appearing in Eq.~(\ref{boltz}) involves a derivative of the total energy density with respect to the visible sector temperature. We differentiate each of the energy densities with respect to $T$ and remembering that $\xi_\alpha=\xi_\alpha(T)$, then Eq.~(\ref{boltz}) can be cast in the form
\begin{align}
\frac{dY_r}{dT}=&-\frac{\mathbb{s}}{H}\Big(\frac{d\rho_v/dT}{4\rho_v-j_v/H}\Big)\mathcal{Q}_r\,.
\label{boltz-r}
\end{align}
It is to be noted from Eq.~(\ref{boltz-r}) that the source term $j_v$ now appears in the Boltzmann equations for the particle yields.
 The formalism given above provides us with a set of coupled Boltzmann equations in temperature and yield which need to be numerically solved to get a full description of the thermal evolution of the hidden sectors. We do this next.

\subsection{Numerical analysis}

In our analysis here we consider the case of self-interacting DM, where we take $m_{\phi_1}/m_D\ll 1$ which will allow us to consider the most recent results on the DM-nucleon spin-independent (SI) cross-section from PandaX-II experiment~\cite{PandaX-II:2021lap} as an experimental constraint. 
In the following we give an analysis of three type of mass hierarchies among
the matter fields of the hidden sectors which are: 
Case 1: $m_{\phi_1}> 2 m_{\phi_2}$, $m_{\phi_2}=m_{\phi_3}$;
Case 2: $m_{\phi_2}> 2 m_{\phi_1}$, $m_{\phi_2}>2m_{\phi_3}$; Case 3: 
$m_{\phi_2}> 2 m_{\phi_1}$, $m_{\phi_3}=0$.
For each of these cases we compute the thermal evolution 
given by $\xi_i=T_i/T$ for the three hidden sectors, and compute the yields.
Specifically the analysis indicates the fraction of the relic density of the dark
matter contributed by each of the dark sectors. We discuss each of these cases in detail below.

\subsection*{Case 1: $m_{\phi_1}>2m_{\phi_2}$ and $m_{\phi_2}=m_{\phi_3}$ }

\begin{figure}[t]
\centering
\includegraphics[width=0.495\textwidth]{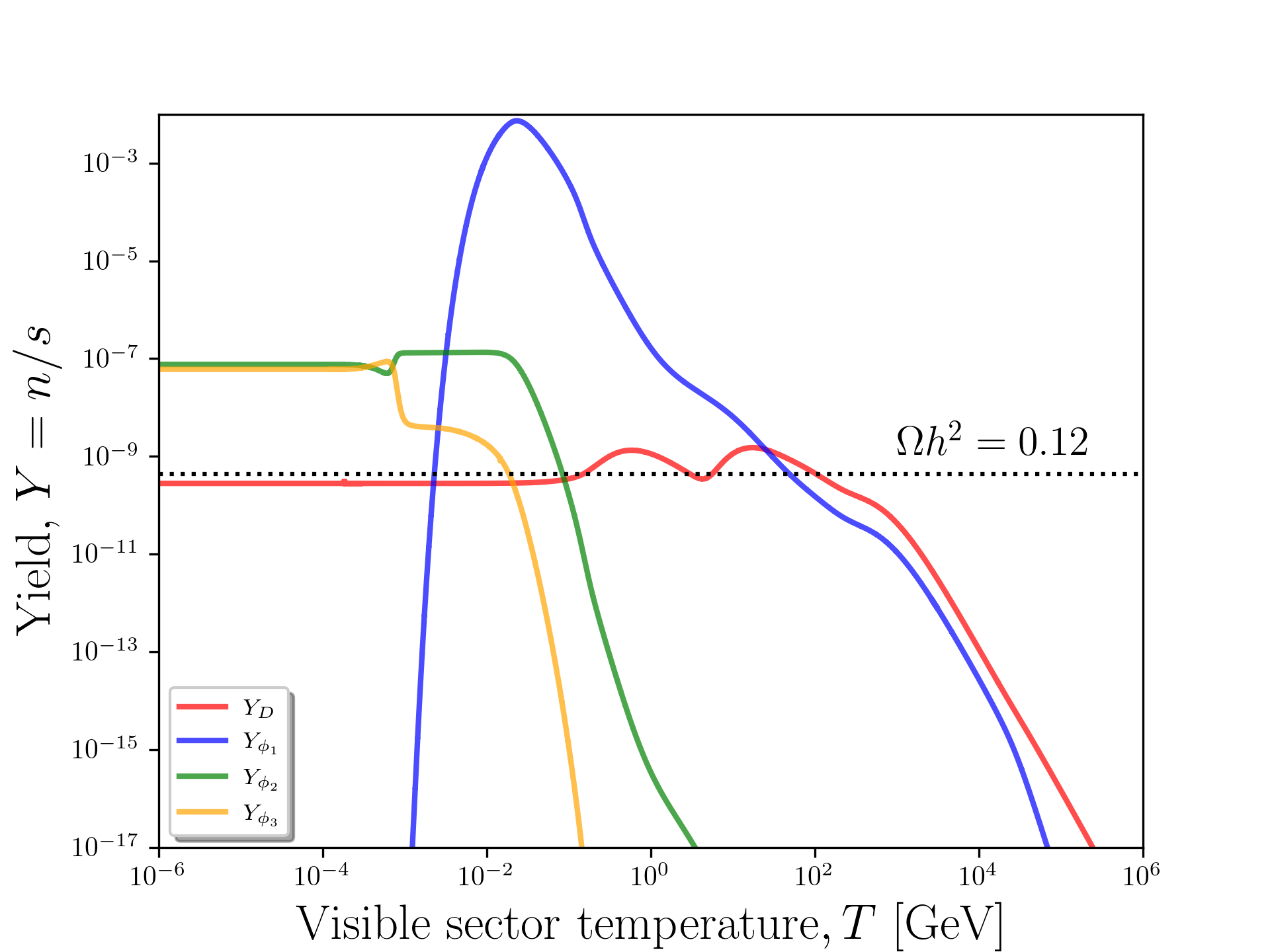}
\includegraphics[width=0.495\textwidth]{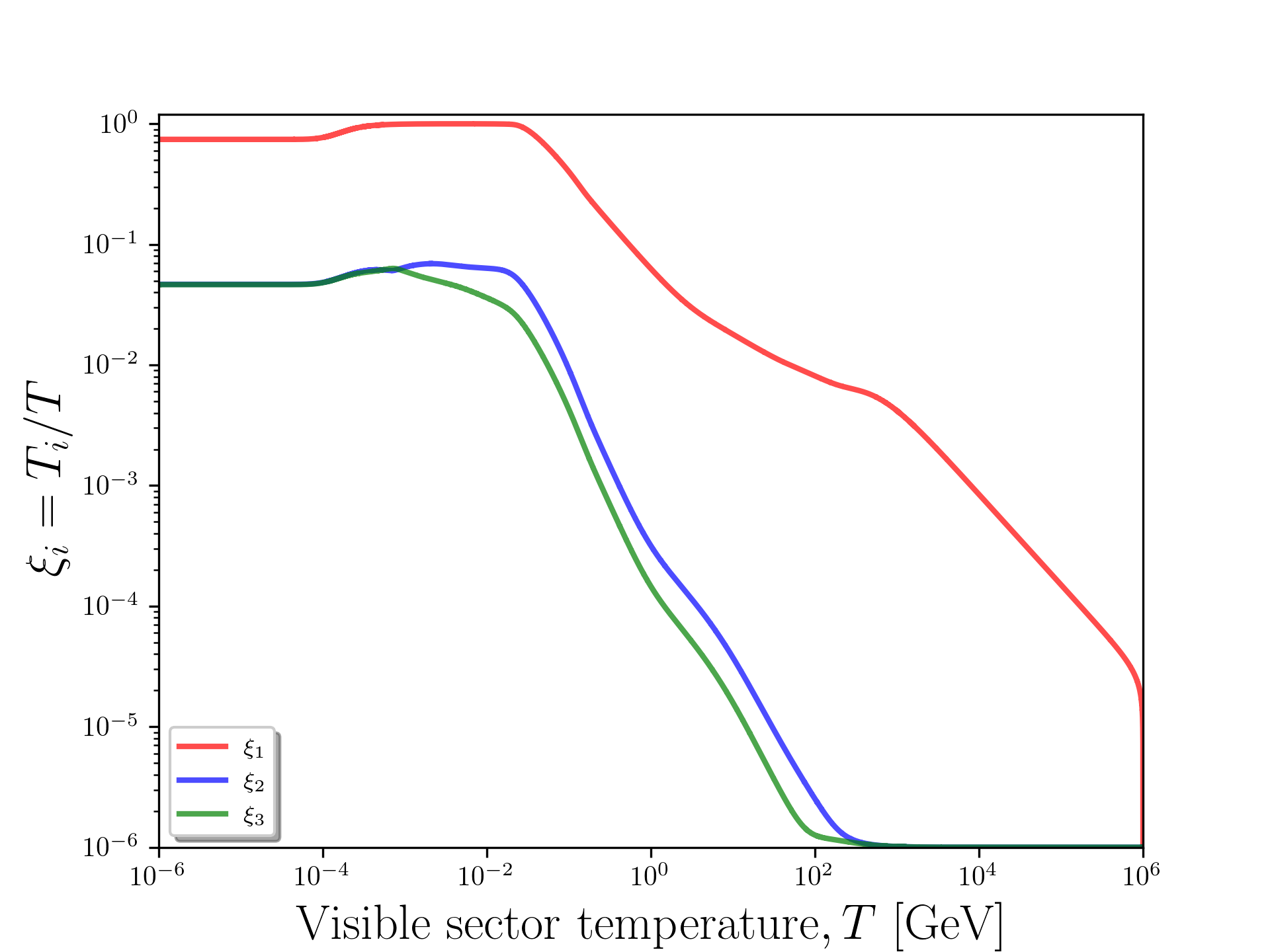}
\caption{The case of three initially cold dark sectors. Left panel: evolution of the yields of the dark fermion $D$ and of the spin zero particles of the dark sector versus the visible sector temperature. Right panel: evolution of the temperature ratios $\xi_i$. The input parameters are: $m_D=1.0$ GeV, $m_{\phi_1}=50$ MeV, $m_{\phi_2}=m_{\phi_3}=1$ MeV, $y_f=1.3\times 10^{-9}$, $y_D=0.053$ and $\kappa_1=\kappa_2=10^{-6}$. Total relic density $\Omega h^2=0.114$, where $\Omega h^2_D=0.072$, $\Omega h^2_{\phi_2}=0.0175$ and $\Omega h^2_{\phi_3}=0.0275$. We note that the kink in $\xi_i$ around $T=10^{-4}$ GeV is due to the variation in the degrees of freedom. Specifically, this is due to neutrino decoupling and $e^+e^-$ annihilation. Similar kinks appear in Figs.~\ref{fig2},~\ref{fig3} and~\ref{fig4}. }
\label{fig1}
\end{figure}

In Fig.~\ref{fig1} we exhibit the particle yields (left panel) and the temperature ratios $\xi_i$ (right panel) as a function of the visible sector temperature. As the temperature $T$ drops, the hidden sectors are gradually populated by the dark particle
species. The process happens sequentially where the first hidden sector 
gets populated first whereas the second and third hidden sectors 
see an increase in the yield much later. This is because the 
first hidden sector is the only one with direct couplings to the visible sector whereas the other two hidden sectors rely on the first hidden sector to get populated.   

Further, the energy densities  of the hidden
sectors depend on the  temperature evolution of the respective sectors. This is illustrated in the right panel of Fig.~\ref{fig1}. 
 Here we notice how $\xi_1$ rises almost immediately starting from large $T$ while $\xi_2$ and $\xi_3$ remain at their lowest values before picking up at $T\sim 30$ GeV. Thus the process of thermalization begins with the first hidden sector
  which then permeates to the second and the third hidden sectors. However,
  while 
 $\xi_1$ reaches 1 the other hidden sectors remain out of thermal equilibrium with $\xi_2,\xi_3\ll 1$ and $\xi_2\neq\xi_3$. For the benchmark considered in Fig.~\ref{fig1}, the relic density is shared among $D$, $\phi_2$ and $\phi_3$ ($\phi_1$ decays back to the SM particles). 
Note that $\phi_2$ can still have a variety of off-shell decays: $\phi_2\to\phi_1^*\phi_1^*\to
  \bar q q \bar q q, \ell^+\ell^- \ell^+\ell^-, \bar q q \ell^+\ell^-, \bar q q \nu \nu,
  \ell^+\ell^- \nu\nu, \nu\nu\nu\nu$. However, if the mass of $\phi_2$ is constrained
  so that $m_{\phi_2}\simeq 1$ MeV, all the decays except for the final state
  $\nu\nu\nu\nu$ will be forbidden.  In this case the decay is extremely suppressed since $\Gamma_{\phi_2}\propto\kappa_1^2 y^4_f$, 
\begin{align}
\Gamma_{\phi_2}&=\frac{\kappa_1^2 y^4_f}{16384\pi^5 m_{\phi_2}}\int_0^{m_{\phi_2}^2}dq_1^2\int_0^{(m_{\phi_2}-q_1)^2}dq_2^2 \nonumber \\
&\hspace{4cm}\times\frac{\lambda^{1/2}(q_1^2,q_2^2,m_{\phi_2}^2)q_1^2 q_2^2}{[(q_1^2-m_{\phi_1}^2)^2+m_{\phi_1}^2\Gamma_{\phi_1}^2][(q_2^2-m_{\phi_1}^2)^2+m_{\phi_1}^2\Gamma_{\phi_1}^2]}.
\end{align} 
Thus we have
 $\tau_{\phi_2}\sim\mathcal{O}(10^{25})$ yrs for the input mentioned in the caption of Fig.~\ref{fig1}.

\begin{figure}[t]
\centering
\includegraphics[width=0.495\textwidth]{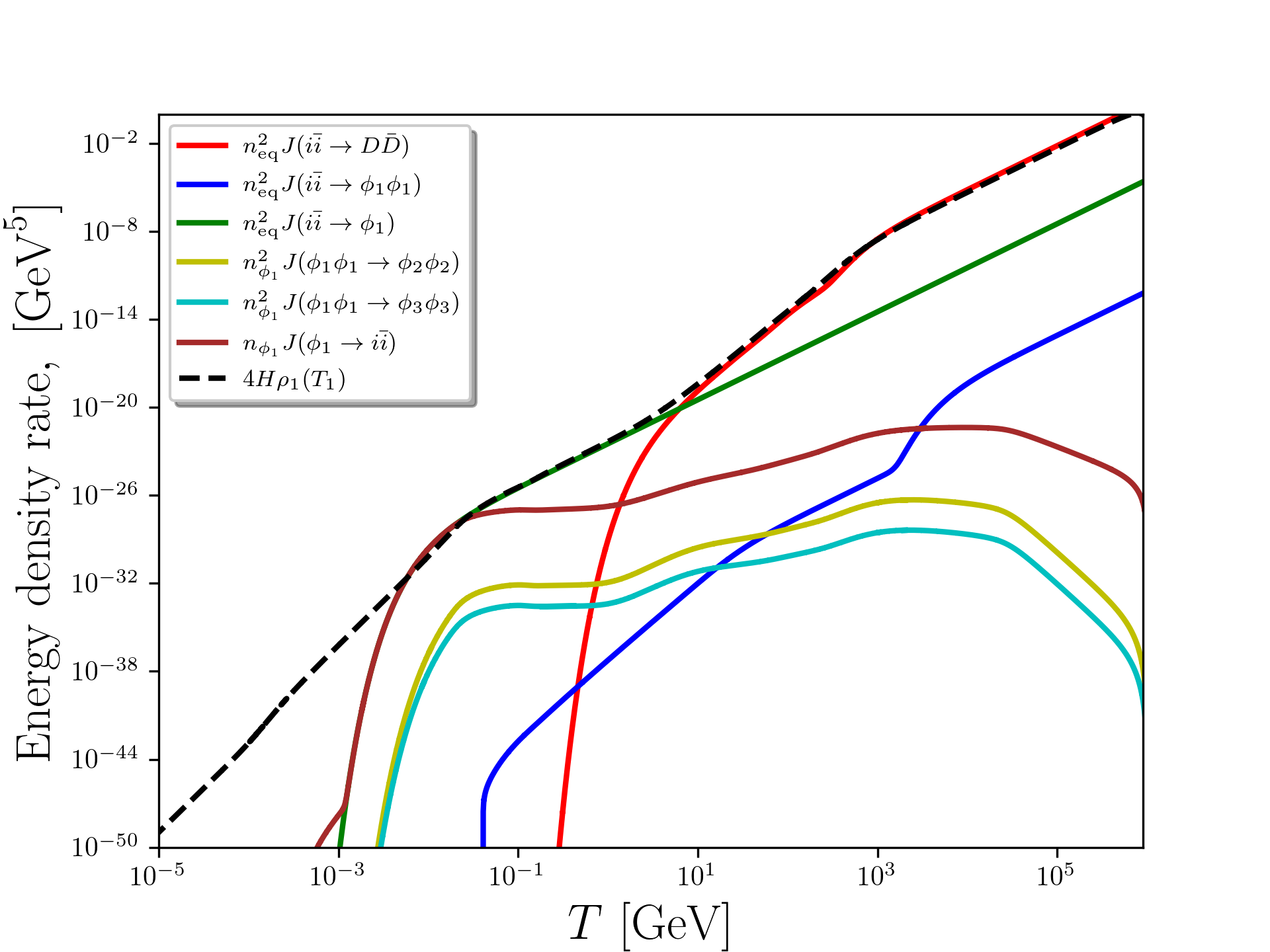}
\includegraphics[width=0.495\textwidth]{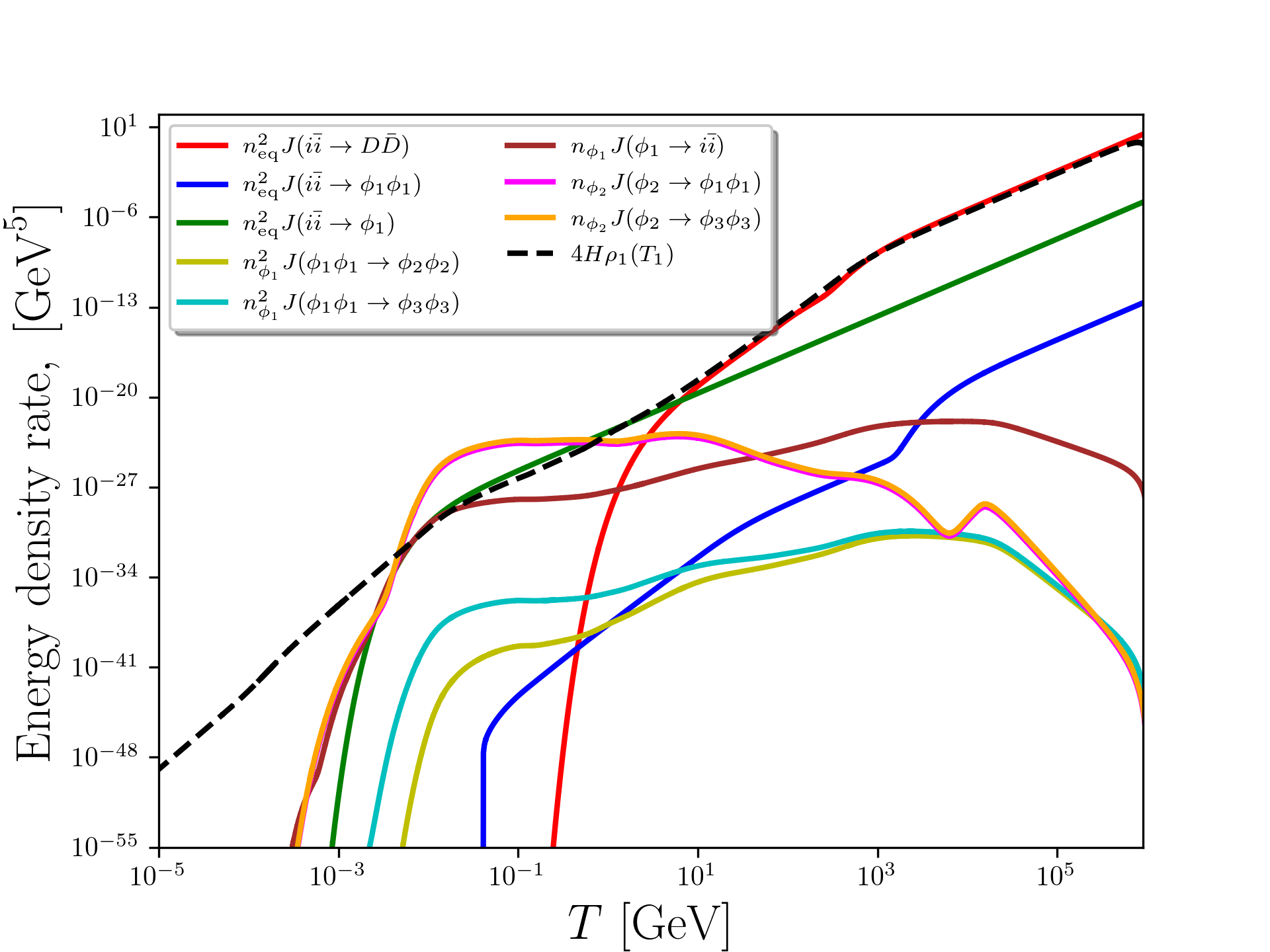}
\caption{Plots of the energy exchange rate between different sectors versus the visible sector temperature $T$. The input for the left panel (Case 1) are same as in Fig.~\ref{fig1} while the input for the right panel (Case 2) are the same as in Fig.~\ref{fig3}. }
\label{fig-J}
\end{figure}

As is apparent from the preceding discussion, the
 thermalization between the first hidden sector and the visible sector is mainly driven by the processes $\phi_1\leftrightarrow i\,\bar{i}$, where $i\,\bar i$ refers to the 
 SM particles. To illustrate this further, we
   plot in Fig.~\ref{fig-J} the energy exchange rate for different processes as well as the energy density dilution term due to the Hubble expansion, $4H\rho_1$. In the left panel of Fig.~\ref{fig-J}, we consider case 1 which corresponds to the analysis of Fig.~\ref{fig1}. We see that all processes are subdominant to the dilution term except for $i\,\bar{i}\to D\bar{D}$ and $\phi_1\leftrightarrow i\,\bar{i}$. Of particular interest are the processes $\phi_1\leftrightarrow i\,\bar{i}$ (green and brown curves) which become comparable in rate near $10^{-2}$ GeV which is the region where the two sectors thermalize as seen in the right panel of Fig.~\ref{fig1}. The rates  slightly overtake $4H\rho_1(T_1)$ for a short period of time before dying out due to Boltzmann suppression of the number density. The right panel of Fig.~\ref{fig-J} pertains to case 2 whose discussion we leave for the next section. 

\begin{figure}[t]
\centering
\includegraphics[width=0.495\textwidth]{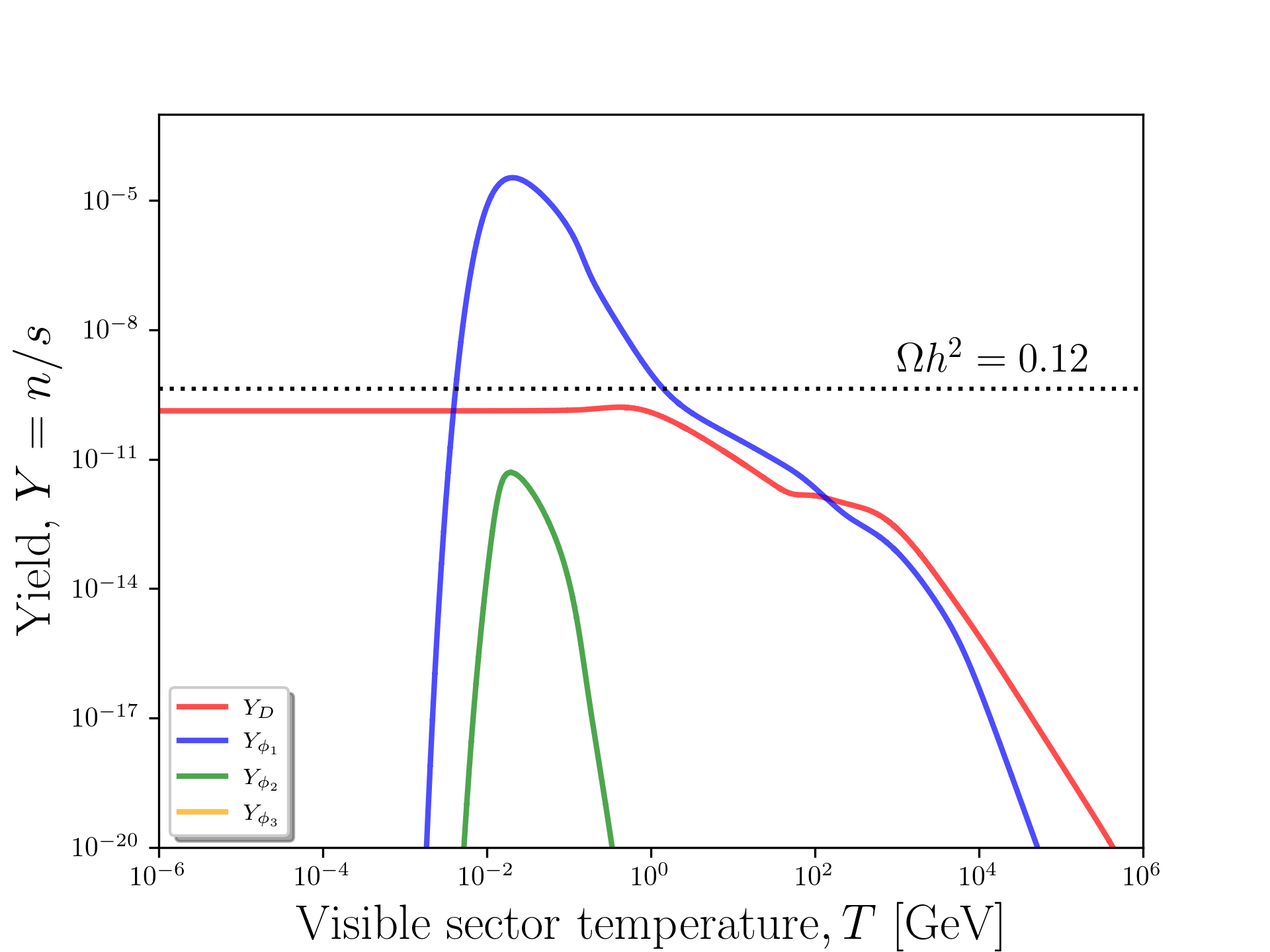}
\includegraphics[width=0.495\textwidth]{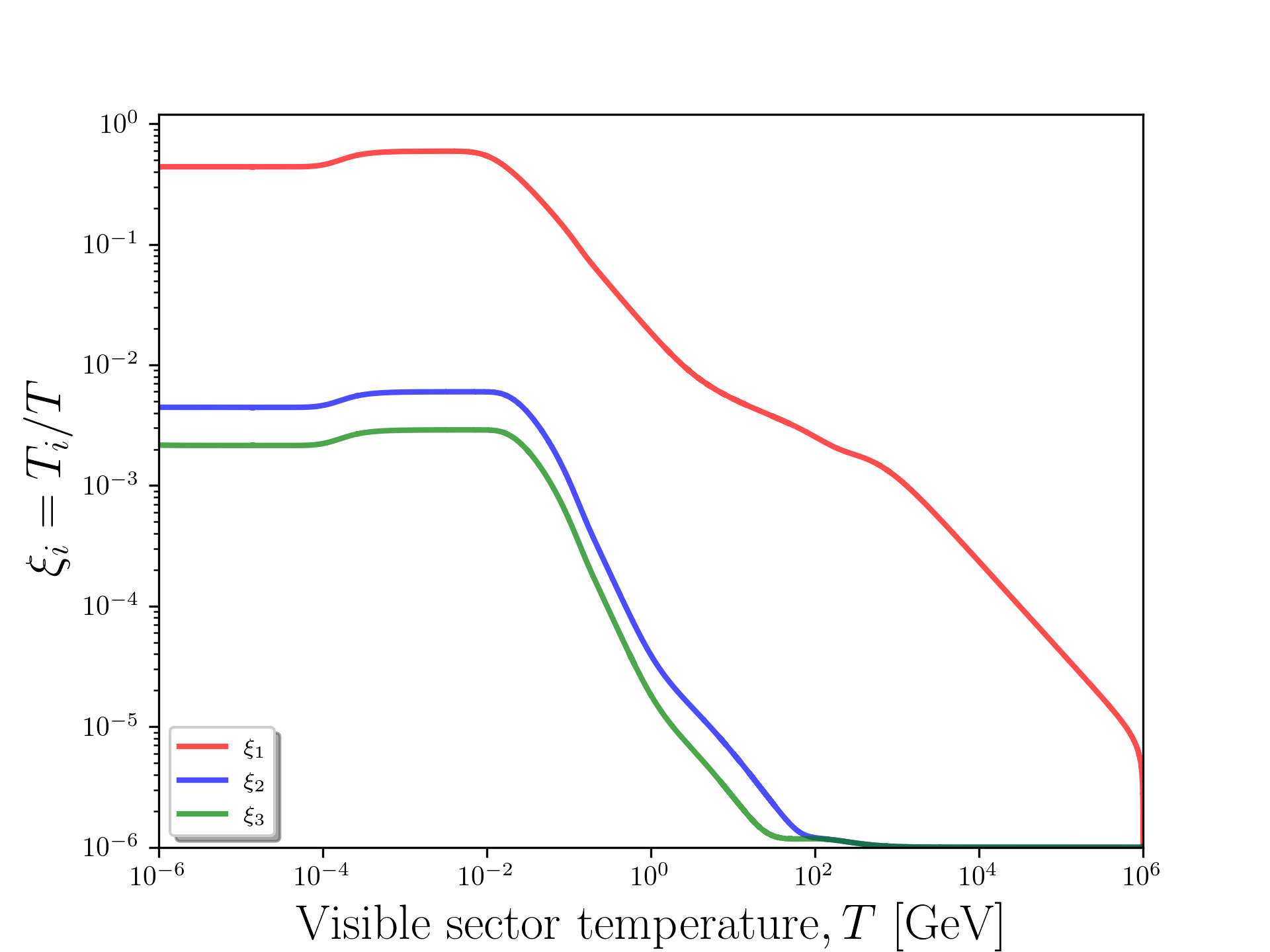}
\caption{Same input as in Fig.~\ref{fig1} but for $y_f=1\times 10^{-10}$. Relic density $\Omega h^2_D=3.6\times 10^{-2}$ and $\Omega h^2_{\phi_2}=\Omega h^2_{\phi_3}\sim 0$.}
\label{fig2}
\end{figure}

Since the visible and the hidden sectors constitute a chain, a change 
 in one coupling will propagate to all sectors. In Fig.~\ref{fig2} we consider a smaller $y_f$ value, i.e. a weaker coupling between the visible and first hidden sector. As expected, the relic density of $D$ becomes smaller and due to a smaller abundance of dark species in the first hidden sector, the yields of $\phi_2$ and $\phi_3$ become negligible (the yield of $\phi_3$ does not appear in the left panel as it is extremely small). Furthermore, the first hidden sector no longer thermalizes with the visible sector as evident from the right panel of Fig.~\ref{fig2}. We also note the change in the evolution of $\xi_2$ and $\xi_3$.

\subsection*{Case 2: $m_{\phi_2}>2m_{\phi_1}$ and $m_{\phi_2}>2m_{\phi_3}$}

\begin{figure}[H]
\centering
\includegraphics[width=0.495\textwidth]{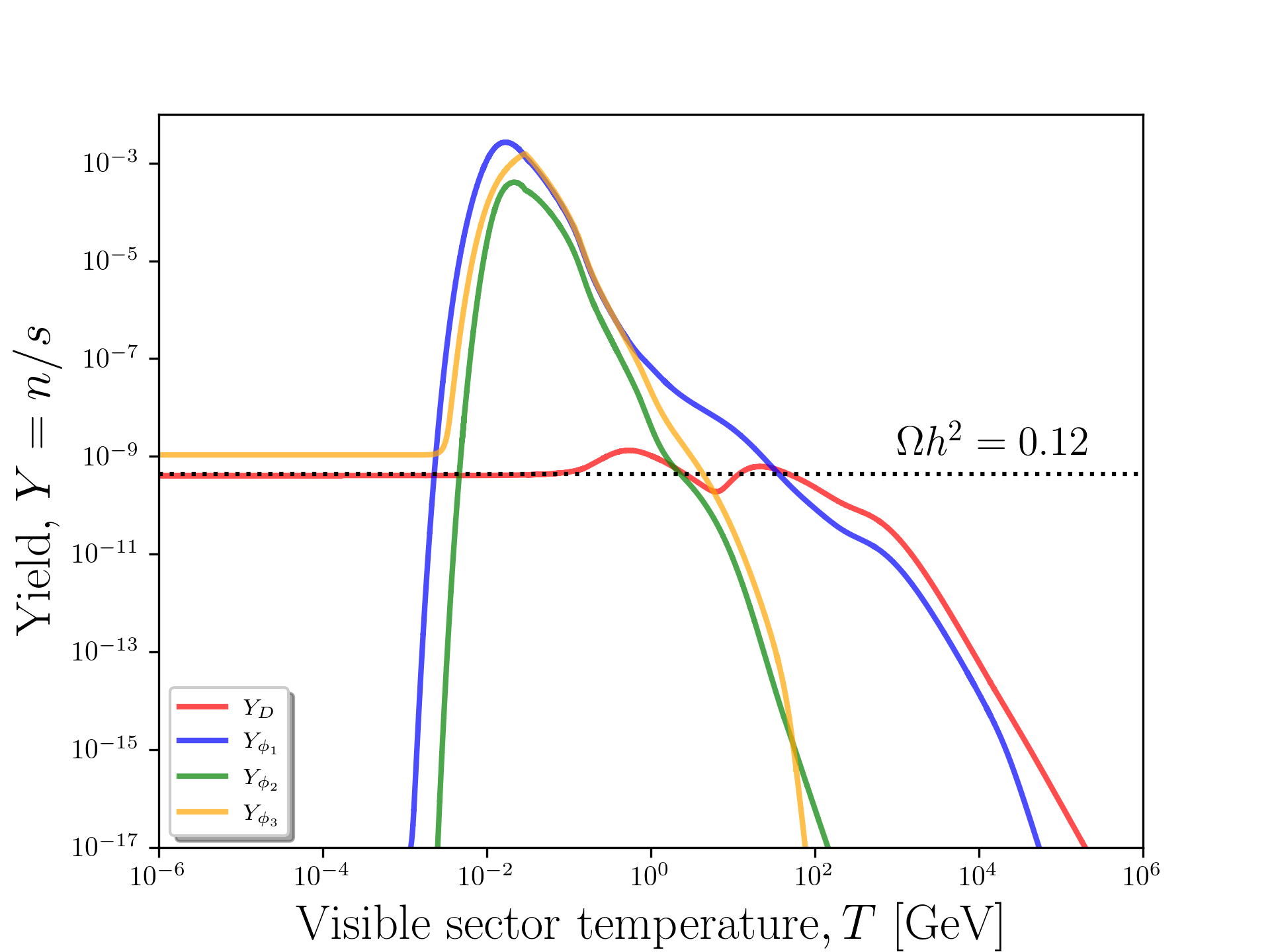}
\includegraphics[width=0.495\textwidth]{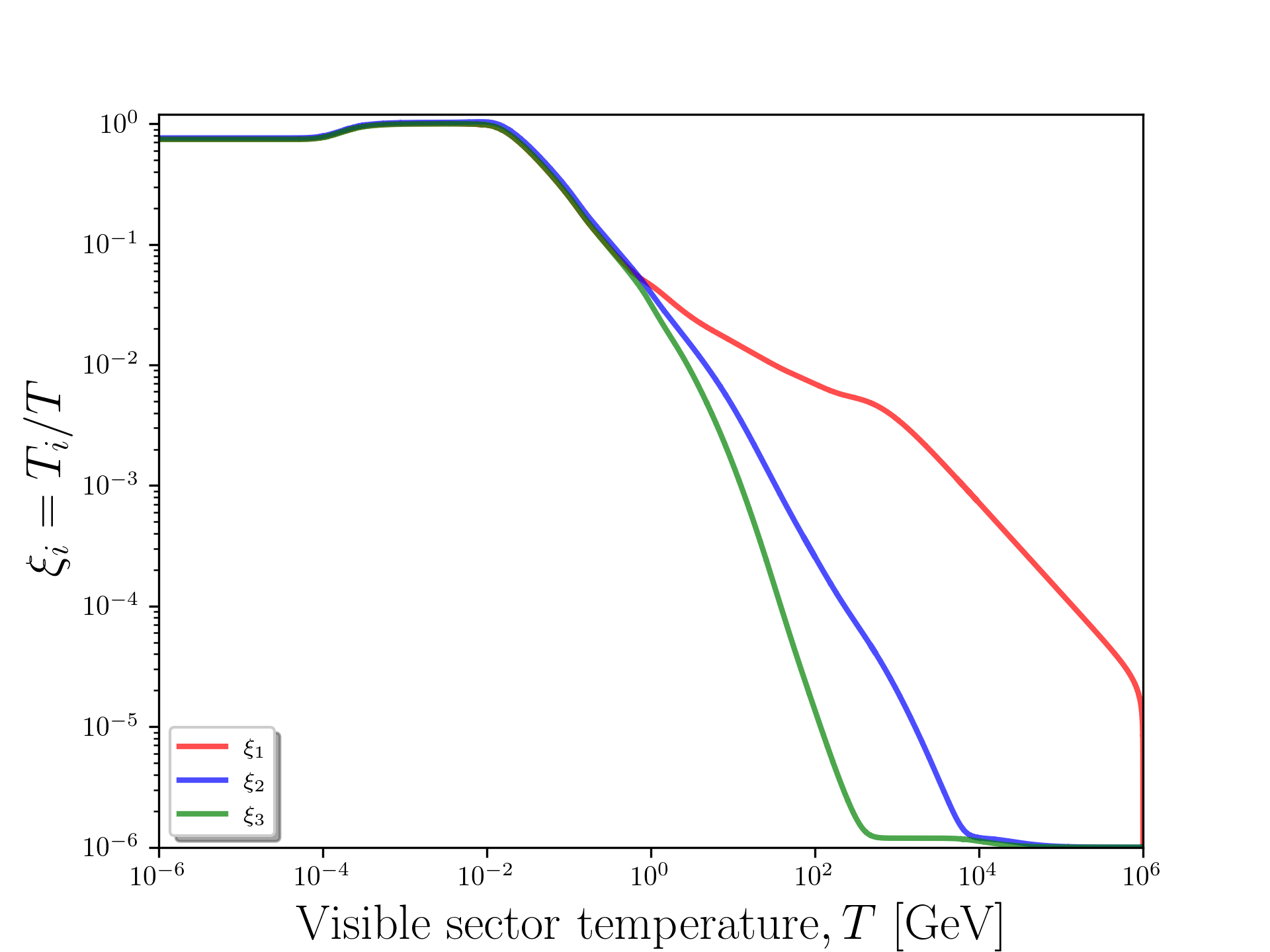}
\caption{The case of three initially cold dark sectors. Left panel: evolution of the yields of the dark fermion $D$ and and of the spin zero particles of the dark sector versus the visible sector temperature. Right panel: evolution of the temperature ratios $\xi_i$. The input parameters are: $m_D=1.0$ GeV, $m_{\phi_1}=50$ MeV, $m_{\phi_2}=105$ MeV, $m_{\phi_3}=45$ MeV, $y_f=1\times 10^{-9}$, $y_D=0.05$,
 $\kappa_1=10^{-7}$ and $\kappa_2=6\times 10^{-6}$. Total relic density is $\Omega h^2=0.1198$, where
 $\Omega h^2_D=0.102$ and $\Omega h^2_{\phi_3}=0.0178$. }
\label{fig3}
\end{figure}

For this mass hierarchy, both decay processes $\phi_2\to\phi_1\phi_1$ and $\phi_2\to\phi_3\phi_3$ are kinematically allowed. We solve the coupled Boltzmann equations for this case and plot in Fig.~\ref{fig3} the yields of the dark species as well as $\xi_i$ against the visible sector temperature.

As one would expect, $\phi_1$ as well as $\phi_2$ completely decay leaving $D$ and $\phi_3$ as the two possible DM components. For the benchmark considered in Fig.~\ref{fig3}, the energy exchange rate due to the processes  $\phi_2\to\phi_3\phi_3$ and $\phi_2\to\phi_1\phi_1$ are comparable and reach values larger than the energy density dilution due to the Hubble expansion (see the magenta and orange curves in the right panel of Fig.~\ref{fig-J}) which results in a rapid thermalization between the three hidden sectors as seen in the right panel of Fig.~\ref{fig3}. Notice that all three sectors remain very much in close thermal contact while also reaching thermal equilibrium with the visible sector for a period of time.

\subsection*{Case 3: $m_{\phi_2}>2m_{\phi_1}$ and $m_{\phi_3}=0$}

\begin{figure}[t]
\centering
\includegraphics[width=0.495\textwidth]{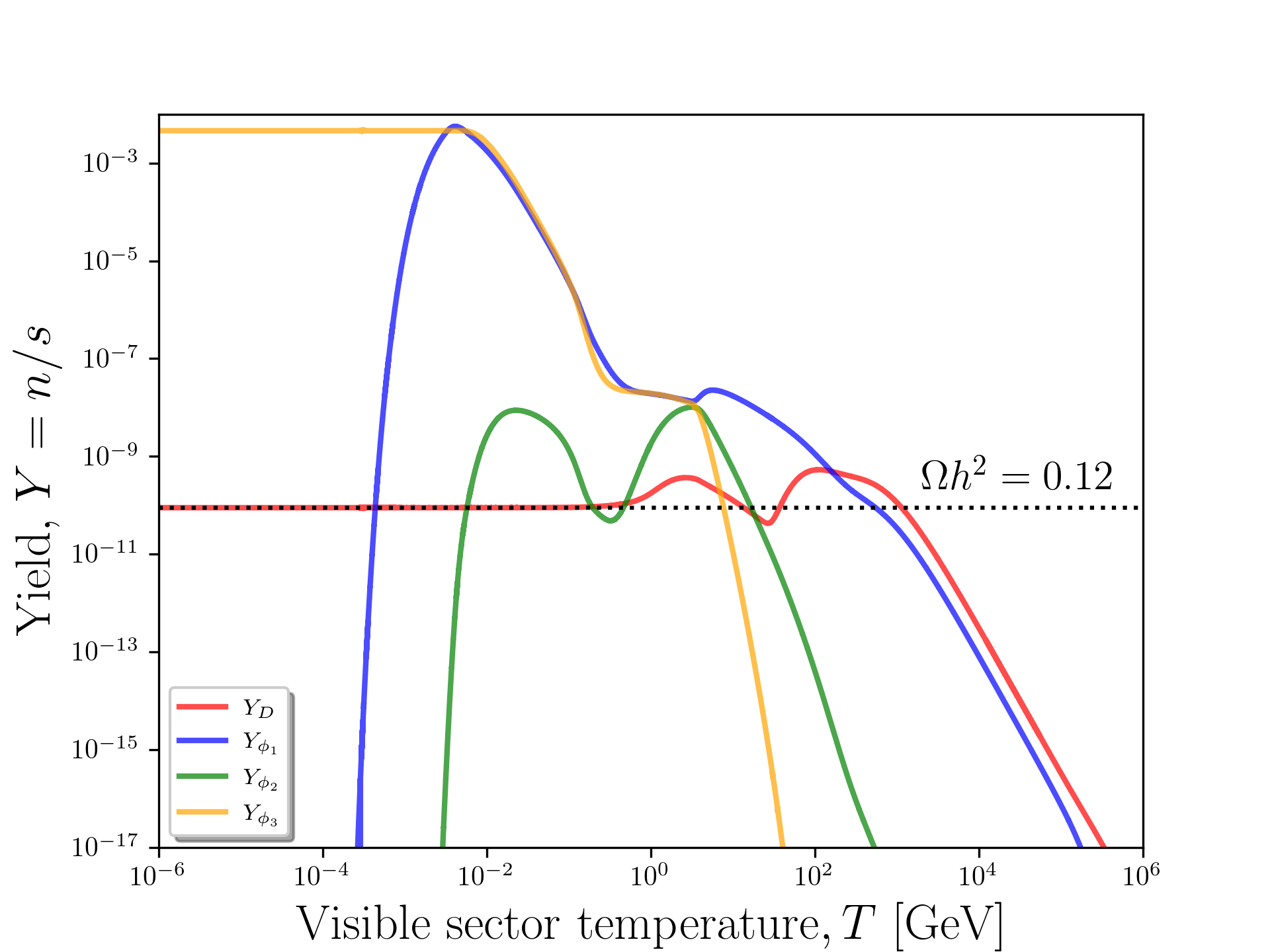}
\includegraphics[width=0.495\textwidth]{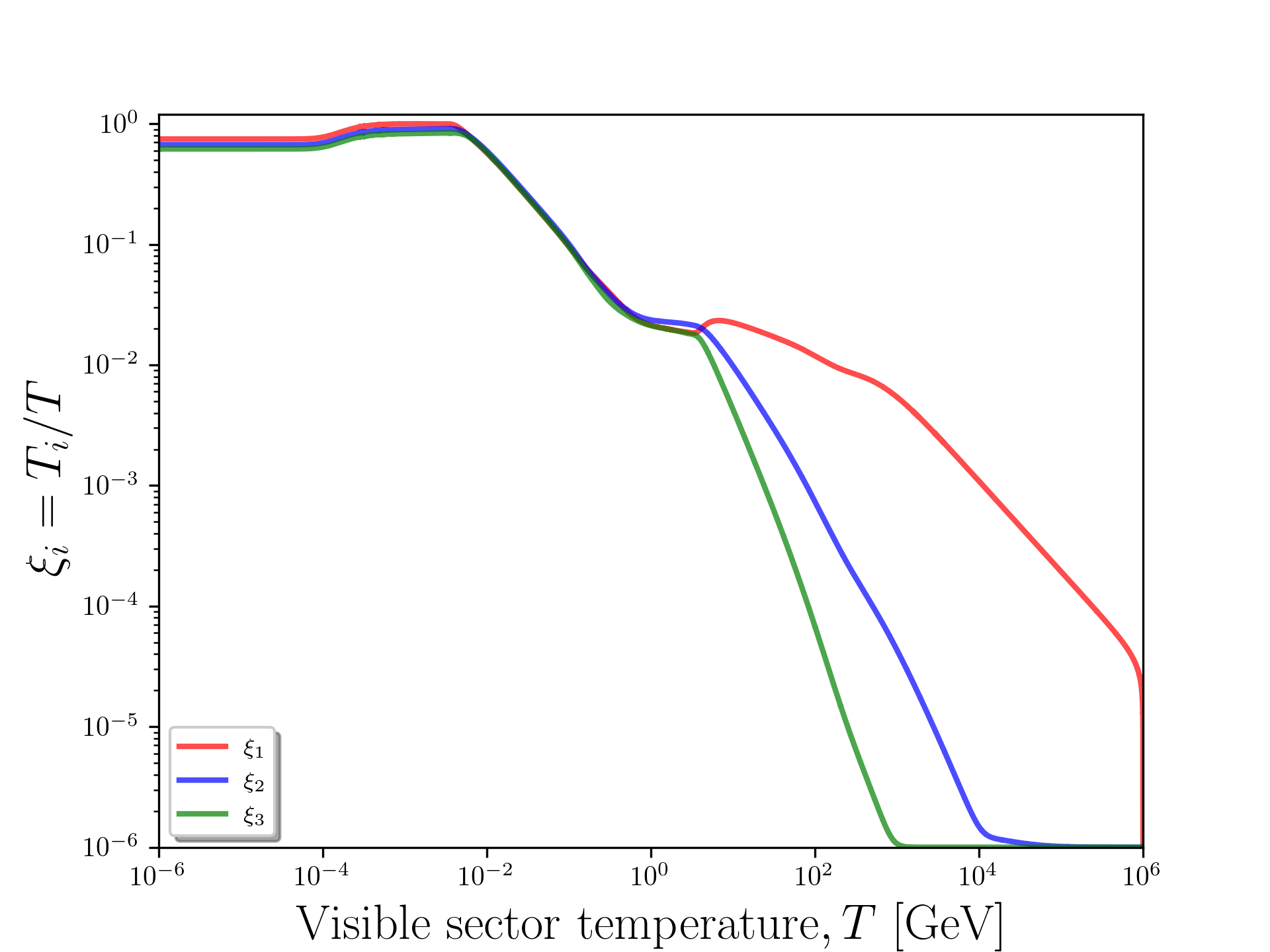}
\caption{The case of three initially cold dark sectors. Left panel: evolution of the yields of the dark fermion $D$ and the scalars of the dark sector versus the visible sector temperature. Right panel: evolution of the temperature ratios $\xi_i$. The input parameters are: $m_D=5.0$ GeV, $m_{\phi_1}=10$ MeV, $m_{\phi_2}=105$ MeV, $m_{\phi_3}=0$, $y_f=6.7\times 10^{-10}$, $y_D=0.17$ and $\kappa_1=\kappa_2=10^{-7}$. Total relic density is $\Omega h^2=\Omega h^2_D=0.123$. As discussed in subsection~\ref{sec5b}, this case leads to
extra relativisitc degrees of freedom at BBN time so that $\Delta N_{\rm eff}^{\rm BBN}=0.323$. }
\label{fig4}
\end{figure}

With the same setup as before we assume now that particle $\phi_3$ is massless. The evolution of the yields of the dark particles as well as the temperature ratios $\xi_i$ are shown in Fig.~\ref{fig4}. The DM relic density is satisfied as seen from the left panel.

Even though $\phi_3$ resides in the third hidden sector, the coupling $y_f$ between the first hidden sector and the visible has an important effect on $\xi_3$ owing to the coupled nature of the system. Smaller values of $y_f$ will lead to smaller $\xi_3$ and hence a lower contribution to $\Delta N_{\rm eff}$. We study this in more detail in the next section.

\section{Implications for observables \label{sec5}}

In this section we discuss the physical consequences of the formalism
given above. In subsection~\ref{sec5a} we discuss the thermal effects of hidden
sectors on the proton-DM scattering cross sections. In subsection~\ref{sec5b}
we discuss the contribution to dark radiation from hidden sectors.

\subsection{Thermal effect of hidden sectors on direct detection of dark matter \label{sec5a}} 

\begin{figure}[t]
\centering
\includegraphics[width=0.495\textwidth]{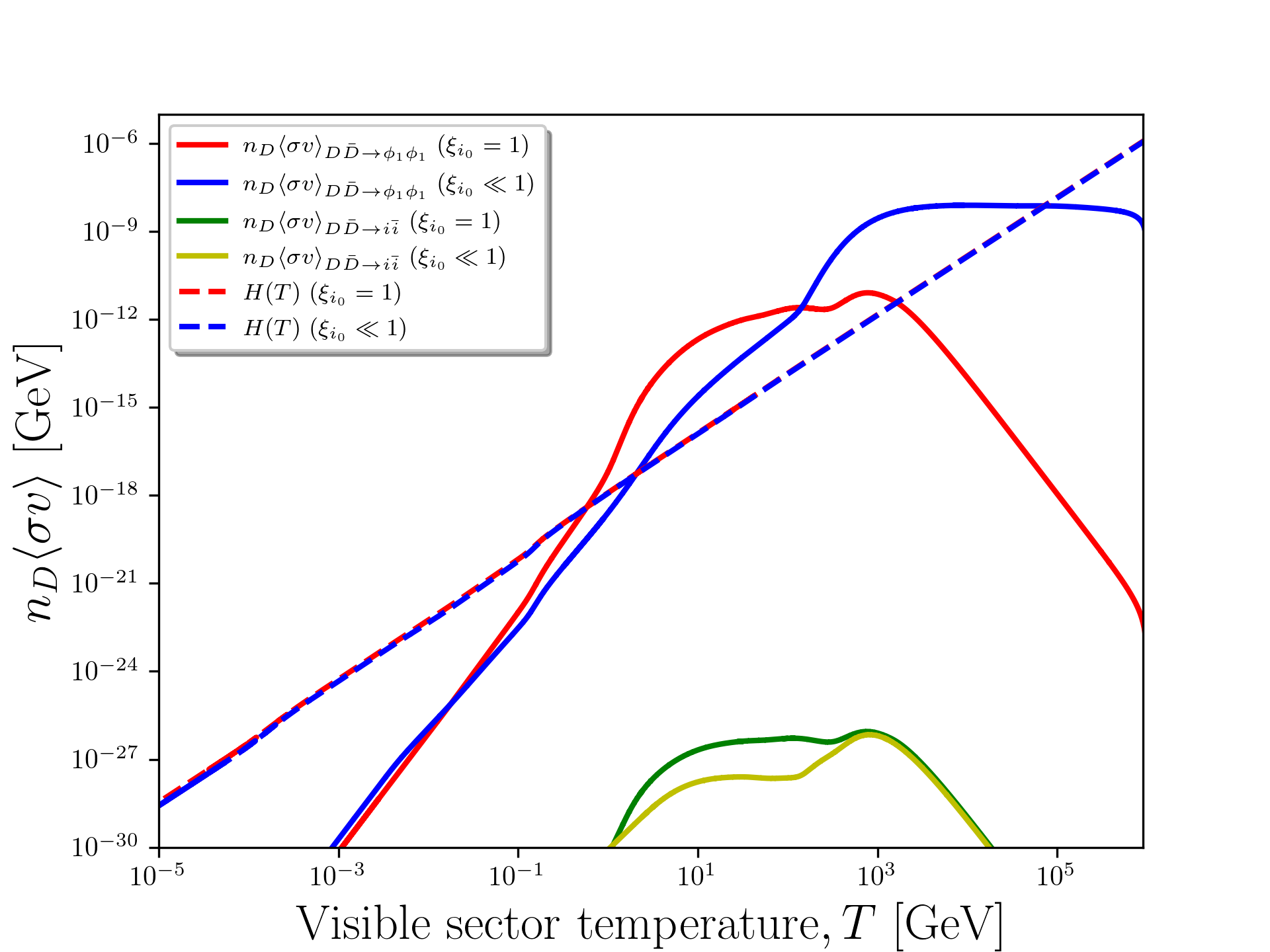}   
\includegraphics[width=0.495\textwidth]{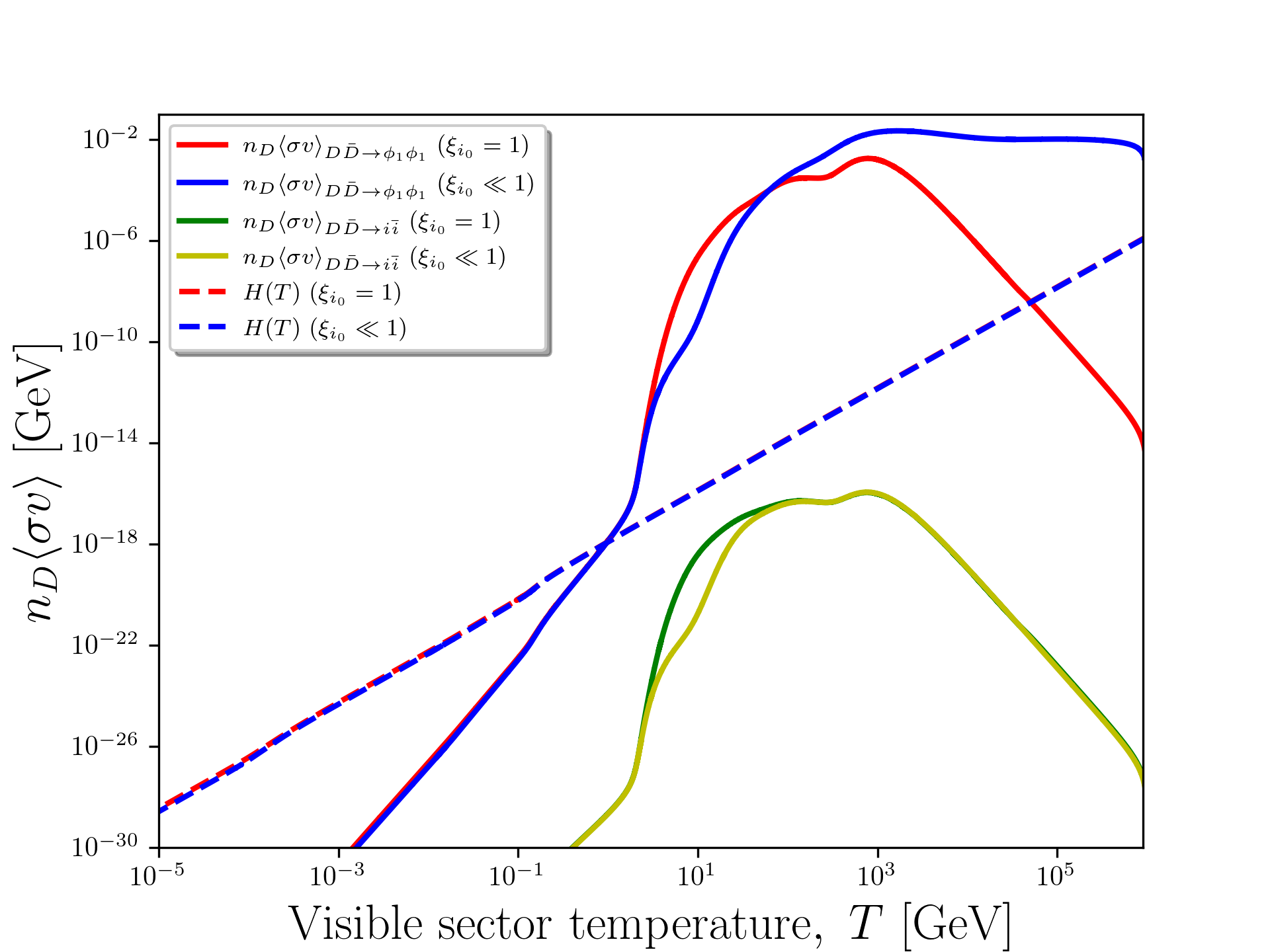}   
\caption{Display of $n_D\langle\sigma v\rangle$ for the processes $D\bar{D}\to\phi_1\phi_1$ and $D\bar{D}\to i\bar{i}$ versus $T$ for two cases: $\xi_{i_0}=1$ and $\xi_{i_0}\neq 1$. Input parameters for the left panel are: $m_D=9.0$ GeV, $m_{\phi_1}=10$ MeV, $m_{\phi_2}=m_{\phi_3}=1$ MeV, $y_f=4.5\times 10^{-10}$, $y_D=0.13$ and $\kappa_1=\kappa_2=5\times 10^{-8}$. Input parameters for the right panel are: $m_D=5.0$ GeV, $m_{\phi_1}=1$ GeV, $m_{\phi_2}=m_{\phi_3}=1$ MeV, $y_f=9.6\times 10^{-6}$, $y_D=0.18$ and $\kappa_1=\kappa_2=1\times 10^{-5}$.}
\label{fig5}
\end{figure}

In this subsection we discuss the effect of a multi-hidden sector model on the determination of the relic density and on the spin-independent DM-proton cross section. 
The first observation to be made is regarding the effect of such a coupled system on the DM relic density. Here we consider case 1 but with $\kappa_1,\kappa_2\sim 10^{-9}$ so that the contributions from $\phi_2$ and $\phi_3$ to the relic density are negligible. We solve the coupled Boltzmann equations for two cases: (A) $\xi_{i_0}=1$ and (B) $\xi_{i_0}\ll 1$.
We note that cases A and B result in different DM relic density such that $(\Omega h^2)_A<(\Omega h^2)_B$. The difference between the two can reach up to 50\% in some cases, especially for $y_f\lesssim 10^{-8}$. To see why this is the case, one needs to examine the main annihilation process responsible for DM depletion. For $m_{\phi_1}\ll m_D$, the DM relic density is dominantly controlled by the annihilation process $D\bar{D}\to\phi_1\phi_1$ whose thermal average depends on the temperature $T_1$. In Fig.~\ref{fig5}, we plot $n_D\langle\sigma v\rangle$ for $D\bar{D}\to\phi_1\phi_1$ versus the visible sector temperature for cases A and B. Also shown (in dashed line) is the Hubble parameter as well as the process $D\bar{D}\to i\,\bar{i}$ for comparison. The left panel of Fig.~\ref{fig5}, which corresponds to a light mediator ($m_{\phi_1}=1$ MeV) and small $y_f$, shows that $D\bar{D}\to\phi_1\phi_1$ is far more dominant than $D\bar{D}\to i\bar{i}$ which can be explained by the relative sizes of the couplings $y_f$ versus $y_D$. Furthermore, the rate of annihilation to $\phi_1\phi_1$ for case B (blue curve) overtakes the expansion rate earlier than for case A (red curve) and consequently decoupling also happens earlier. Thus for case A, the annihilation process continues for a longer time leading to a smaller relic density. This is in contrast to the case of a heavier mediator ($m_{\phi_1}=1$ GeV) and a larger $y_f$ as shown in the right panel of Fig.~\ref{fig5} where one can see that $D\bar{D}\to\phi_1\phi_1$ decouples at the same instant for both cases A and B and thus the final yield is insensitive to whether the two sectors were initially at the same temperature or not. 

\begin{figure}[t]
\centering
\includegraphics[width=0.495\textwidth]{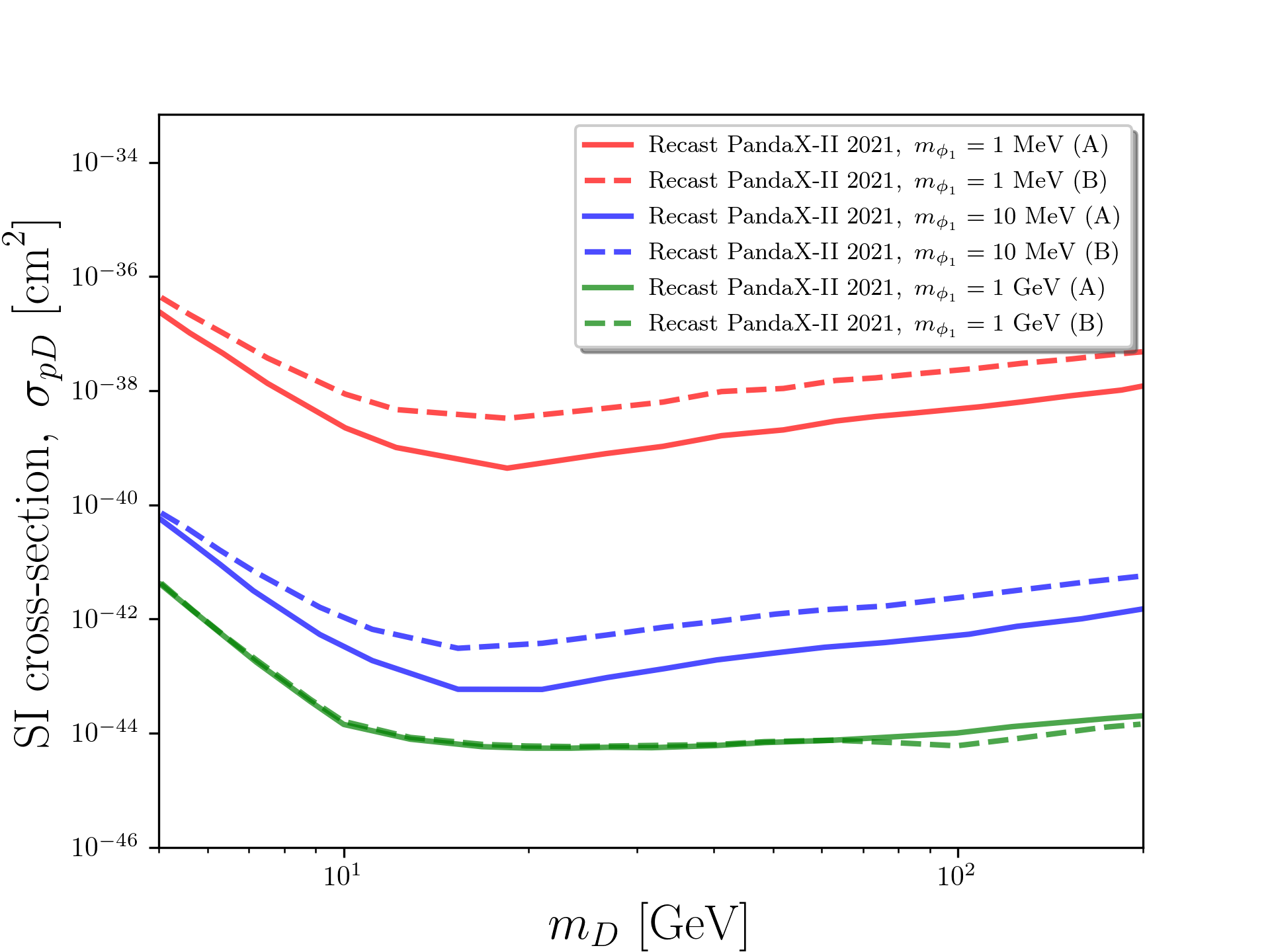}
\includegraphics[width=0.495\textwidth]{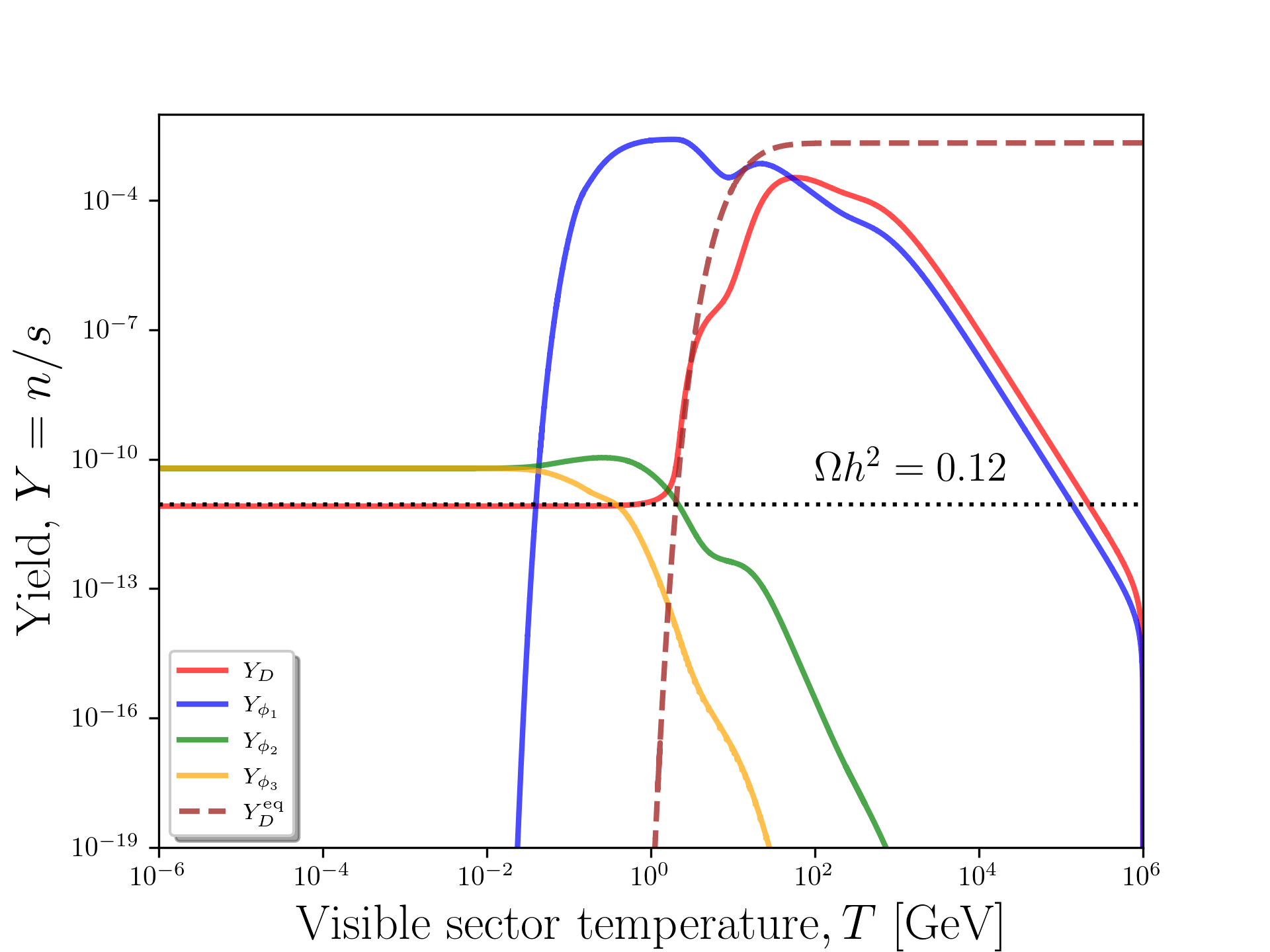}
\caption{Left panel: A recast of the exclusion limits of the spin-independent proton-DM cross section,
$\sigma_{pD}$, given by the PandaX-II Collaboration~\cite{PandaX-II:2021lap} for our model with 
 two sets of initial hidden sector constraints, i.e., case A
with $\xi_{i0}=1$ (solid lines) and case B with $\xi_{i0}<<1$ (dashed lines)
as a function of $m_D$ for  three different values of the pseudo-scalar
$\phi_1$ mass, i.e., $m_{\phi_1}$=(1 MeV (red lines), 10 MeV (blue lines)
and 1 GeV (green lines)).
The coupling $y_D$ for each point is chosen to give the correct DM relic density. Right panel: a plot of the yield of the dark particle species with the input: $m_D=48$ GeV, $m_{\phi_1}=1$ GeV, $m_{\phi_2}=m_{\phi_3}=1$ MeV, $y_f=10^{-7}$, $y_D=0.55$ and $\kappa_1=\kappa_2=1\times 10^{-5}$.}
\label{fig6}
\end{figure}

To illustrate more concretely the effect of hidden sectors on observables
in the visible sector, 
we use the latest constraints on the spin-independent (SI) DM-proton scattering cross-section  from the PandaX-II collaboration. The collaboration presented their limits based on a secluded dark sector model of ref.~\cite{Huo:2017vef} which considers a thermal DM distribution in the early universe. Using those limits, we recast the PandaX-II results to our model parameters and display the limits in the left panel of Fig.~\ref{fig6} (solid lines). 
The analysis which corresponds to the solid lines is for 
case A where we assume that the sectors have the same initial temperature. The dashed lines correspond to case B where the hidden sectors are initially much colder than the visible sector. 
 We note
 that for 1 MeV and 10 MeV mediators the limits have moved up, i.e., have become more relaxed while for a 1 GeV mediator there is virtually no effect.
 
  As  discussed earlier, sectors at different temperatures lead to a larger relic density and so to reduce the value of the latter, one increases the coupling $y_D$ which leads to a larger SI cross-section. Thus the limits move up as shown in the left panel of Fig.~\ref{fig6} (dashed lines). However, this seems not to be the case for a 1 GeV mediator. Also as discussed earlier, the main annihilation process decouples at the same time for cases A and B which results in no change in the relic density. This is due to the fact that in this case a larger $y_f$ is required for those limits ($y_f\gtrsim 10^{-7}$) which takes us closer to a pure freeze-out case as the one used by PandaX-II to produce those limits. Thus as the coupling $y_f$ increases, it gradually phases out the freeze-in mechanism~\cite{Hall:2009bx} and we begin to agree with the limits presented in ref.~\cite{PandaX-II:2021lap}. This can be clearly seen in the right panel of Fig.~\ref{fig6} where the yield of $D$ (red curve) reaches close to its equilibrium value (dashed brown curve) and then undergoes a freeze-out in a manner different from what appears for the cases of smaller $y_f$ (see Figs.~\ref{fig1} and \ref{fig2}$-$\ref{fig4}).  

\subsection{Dark radiation from hidden sectors \label{sec5b}}

Hidden sectors in general contain extra relativistic degrees of freedom which
will contribute to the so-called dark radiation which is constrained by BBN and
CMB data. Thus, for example, for case 3, we have a massless spin zero field $\phi_3$ which will remain in the spectrum and  contribute  to the relativistic degrees of freedom in the universe as dark radiation.
  Such a contribution over the existing neutrino species is determined by $\Delta N_{\rm eff}$, so that
\begin{equation}
\Delta N_{\rm eff}=\frac{8}{7}\left(\frac{11}{4}\right)^{4/3}\frac{\rho_{\rm DS}}{\rho_\gamma},
\label{dneff}
\end{equation} 
where the energy density of the dark sector is $\rho_{\rm DS}=\rho_{D}(T_1)+\rho_{\phi_3}(T_3)$ and $\rho_\gamma$ is the photon energy density. Based on the latest estimate of the effective number of neutrino species, $N_{\rm eff}$, from BBN~\cite{Pitrou:2018cgg}
\begin{equation}
N_{\rm eff}=2.88\pm 0.27,    
\label{pitrou}
\end{equation}
where one can deduce a $2\sigma$ upper bound on $\Delta N_{\rm eff}$, so that $\Delta N_{\rm eff}<0.42$. If $\phi_3$ is produced efficiently and reaches an equilibrium distribution, then $\rho_{\phi_3}\propto T_3^4$ and from Eq.~(\ref{dneff}) we have $\Delta N_{\rm eff}\sim (T_3/T)^4=\xi_3^4$. So of particular importance to the determination of $\Delta N_{\rm eff}$ is the evolution of $\xi_3$ which is tightly coupled to the rest of the evolution parameters. In Fig.~\ref{fig4} we display the evolution of the yields and $\xi_i$ for the case of a massless $\phi_3$. For the set of assumed couplings, we have $\Delta N_{\rm eff}^{\rm BBN}=0.323$ consistent with the BBN bound. 

\begin{figure}[t]
\centering
\includegraphics[width=0.65\textwidth]{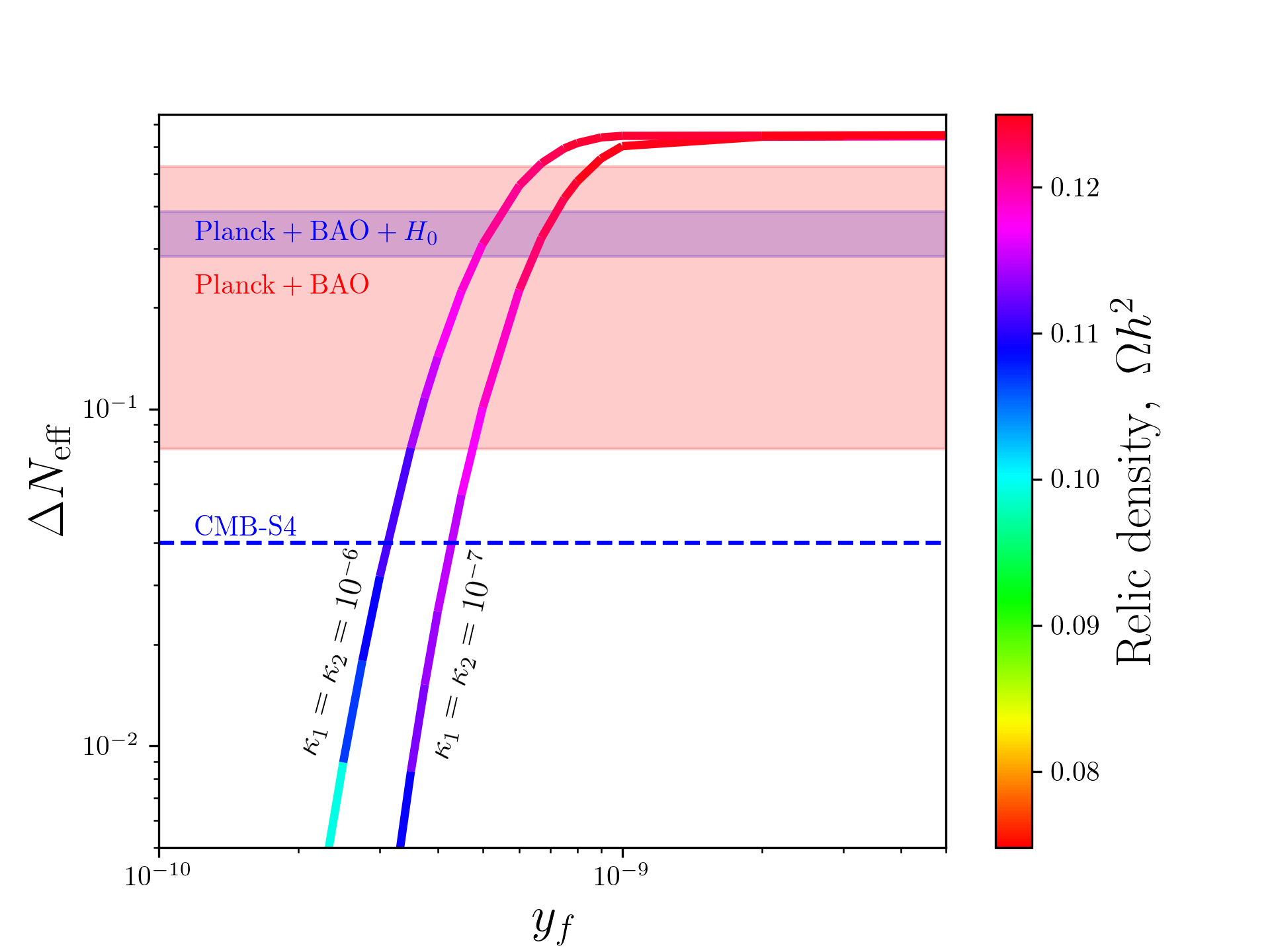}
\caption{A plot of $\Delta N_{\rm eff}$ for the case of massless $\phi_3$ as a function of $y_f$ for sets of values of $\kappa_1$ and $\kappa_2$. The color gradient of the curves reflects the DM relic density. Other input parameters are same as in Fig.~\ref{fig4}. }
\label{fig7}
\end{figure}

As noted, the
 constraint given in Eq.~(\ref{pitrou})
comes from BBN while other estimates of $N_{\rm eff}$ rely on the power spectrum data and CMB lensing from Planck and Baryon Acoustic Oscillations (BAO)~\cite{Planck:2018vyg,Planck:2018nkj}. Further, local measurements of the Hubble parameter $H_0$~\cite{Riess:2021jrx,Riess:2018uxu,Riess:2019cxk} show a significant deviation from estimates of $H_0$ at earlier times (see ref.~\cite{Abdalla:2022yfr} for a review of tensions in cosmology and ways to alleviate them and ref.~\cite{Aboubrahim:2022gjb} for a cosmologically consistent particle physics model aimed at alleviating the Hubble tension). A fit of the cosmological parameters to the CMB, BAO and local $H_0$ data points to a larger $\Delta N_{\rm eff}$ at the surface of last-scattering. The corresponding ranges for $\Delta N_{\rm eff}$ based on these analyses are shown in Fig.~\ref{fig7} as red and blue bands. The future CMB-S4 experiment~\cite{CMB-S4:2016ple,Abazajian:2022nyh} is expected to reach a much better sensitivity in determining $N_{\rm eff}$ which will put stronger constraints on models predicting extra light relics. The projected sensitivity of CMB-S4 is shown as the dashed blue line in Fig.~\ref{fig7}.    

The curves in Fig.~\ref{fig7} show the variation in $\Delta N_{\rm eff}$ as a function of the coupling $y_f$ for two sets of $\kappa_1,\kappa_2$. Since the hidden sectors form a chain with the visible sector then a change in $y_f$ will eventually propagate to the third hidden sector thus affecting the energy density of $\phi_3$ which is a function of the temperature $T_3$ (here enters the dependence on $\xi_3$). Larger $y_f$ leads to larger contributions to $\Delta N_{\rm eff}$ until it saturates for $y_f\gtrsim 10^{-9}$ for the particular benchmark under consideration. Furthermore, larger values of $\kappa_1$ and $\kappa_2$ lead to a stronger coupling between the hidden sectors which eventually leads to an enhanced contribution to $\Delta N_{\rm eff}$ as evident when comparing the two  curves in Fig.~\ref{fig7}. The color gradient of the curves shows the change in the relic density as $y_f$ varies. One can see that in the range of interest $\Omega h^2\leq 0.12$. Based on our model, CMB-S4 can probe feeble couplings between the visible and hidden sectors down to $\mathcal{O}(10^{-10})$. Deviations from $N_{\rm eff}^{\rm SM}$ which are $\mathcal{O}(10^{-2})$ can point to relativistic relics that reside in hidden sectors at temperatures much lower than the visible sector temperature.

\section{Conclusion}\label{sec:conc}

In a variety of supergravity, string and brane models one encounters 
hidden sectors which although neutral under the Standard Model gauge
group can interact feebly with the visible sector via a variety of portals.
 In this work we have extended previous works where the visible sector
 was coupled to one or two hidden sectors to the case where an arbitrary number of hidden sectors
 are sequentially coupled to the visible sector. Thus the visible 
 sector couples only to hidden sector 1 and the latter couples to hidden sector 2, and so on, i.e., in general the  $i$-th hidden
 sector couples to hidden sectors  ($i\pm 1$). In general the visible
 and the hidden sectors will be in different heat baths, each with their
 own temperature.  Of prime interest is to obtain the evolution equations
 for the ratio of temperatures in different heat baths after one has chosen
 a given temperature to serve as the clock. In the work here we use the 
 temperature $T$ of the visible sector  as the clock and 
 obtain the evolution equations for the ratios $T_i/T, i=1,2,\cdots, n$.
 A detailed analysis for the case of the visible sector coupled to three 
 hidden sectors is given with an application to a simple model with scalar portals. We solve a set of coupled Boltzmann equations for particle yields and the temperature ratios $\xi_i$ to trace the evolution of the dark species' population as well as the temperatures in each of the hidden sectors. We've shown that the presence of a chain of hidden sectors affects the DM relic density which comes out to be larger than one would get assuming all sectors are at the same temperature. A direct consequence of this observation is a modification of the recent PandaX-II limits on the SI DM-proton scattering cross sections for the case of self-interacting DM. Also we have seen that the hidden sectors thermalize with each other once the decay channels across the sectors are active. Further, assuming that one of the spin zero particles is massless, contributions to $\Delta N_{\rm eff}$ become an important constraint from cosmology. Weaker couplings between the different sectors lead to a smaller dark species' energy density and temperature ratios $\xi_i$. We expect the future CMB-S4 experiment to probe $\Delta N_{\rm eff}$ down to $\mathcal{O}(10^{-2})$ which would constraint our model in the very feeble coupling regime. We note that applications of the formalism developed here to other portals
 and to larger values of $n$ beyond $n=3$ and possibly to explore the region
 of large $n$ should be of interest for future investigations.
 Finally, as noted in section~\ref{sec:model}, a chain of $n$ hidden sectors appears in a broad class  of beyond the Standard Model physics scenarios such as moose/quiver theories. Thus an analysis of the type discussed above may find application in this broad class of models.

\section*{\small Acknowledgments}

The research of AA was supported by the BMBF under contract 05P21PMCAA and by the DFG through the Research Training Network 2149 ``Strong and Weak Interactions -- from Hadrons to Dark matter", while the research of PN was supported in part by the United States NSF Grant PHY-1913328.

\appendix

\section{Mass and temperature dependence of $g_{\rm eff}$ and $h_{\rm eff}$}  \label{app:A}

The total effective number of energy density and entropy degrees of freedom in a sector containing a Dirac fermion and a boson are given by
\begin{align}
g_{\rm eff}&= g^{b}_{\rm eff} +\frac{7}{8}  g^f_{\rm eff},~~~\text{and}~~~h_{\rm eff}= h^{b}_{\rm eff} + \frac{7}{8} h^f_{\rm eff},
\end{align}
where the superscript $b~(f)$  indicates bosonic (fermionic) degrees of freedom 
where at temperature $T$, 
 $g^b_{\rm eff}, h^b_{\rm eff}$   and  $g^f_{\rm eff}, h^f_{\rm eff}$  are given by~\cite{Hindmarsh:2005ix} 
\begin{equation}
\begin{aligned}
g^{b}_{\rm eff}& = \frac{15 d_b}{\pi^4} \int_{x_{b}}^{\infty} \frac{\sqrt{x^2-x_{b}^2} }{e^x-1 } x^2 dx,~~~\text{and}~~~h^{b}_{\rm eff}= \frac{15 d_b}{4\pi^4} \int_{x_{b}}^{\infty} \frac{\sqrt{x^2-x_{b}^2} }{e^x-1 } 
(4x^2-x_{b}^2) dx, \\
g^{f}_{\rm eff}& = \frac{15 d_f}{\pi^4} \int_{x_{f}}^{\infty} \frac{\sqrt{x^2-x_{f}^2} }{e^x+1 } x^2 dx,~~~\text{and}~~~h^{f}_{\rm eff}= \frac{15d_f}{4\pi^4} \int_{x_{f}}^{\infty} \frac{\sqrt{x^2-x_{f}^2} }{e^x+1 } 
(4x^2-x_f^2) dx.
\end{aligned}
\label{hdof}
\end{equation}
where $d_b~(d_f)$ are the intrinsic degrees of freedom for a bosonic (fermionic) field, so
that for a massive spin 1, $d_b=3$ and for a Dirac fermion $d_f=4$.
Further,  $x_{b}$ and $x_f$ are defined so that $x_{b}= m_{b}/T$ and $x_f= m_f/T$. 
The  limit  $x_{b}\to 0$  gives $g^{b}_{\rm eff}= h^{b}_{\rm eff}\to d_b$ and the limit $x_f\to 0$ gives $g^{f}_{\rm eff}= h^{f}_{\rm eff}\to d_f$.

\section{Collision and source terms}\label{app:B}

In this appendix we list the collision terms appearing in Eq.~(\ref{boltz-r}) and the source terms appearing in Eqs.~(\ref{xi1})$-$(\ref{xi3}) for particles $D$, $\phi_1$, $\phi_2$ and $\phi_3$.
The respective collision terms are given by
\begin{align}
\mathcal{Q}_D=&\frac{1}{2}\langle\sigma v\rangle_{D\bar{D}\to f\bar{f}}(T)(Y_D^{\rm eq}(T)^2-Y_D^2)-\frac{1}{2}\langle\sigma v\rangle_{D\bar{D}\to\phi_1\phi_1}(T_1)\left(Y^2_D-Y_D^{\rm eq}(T_1)^2\frac{Y^2_{\phi_1}}{Y^{\rm eq}_{\phi_1}(T_1)^2}\right),
\label{CD}
\end{align}
\begin{align}
\label{yp1}
\mathcal{Q}_{\phi_1}=&\langle\sigma v\rangle_{\phi_1\phi_1\to f\bar{f}}(T)(Y_{\phi_1}^{\rm eq}(T)^2-Y_{\phi_1}^2)+\frac{1}{2}\langle\sigma v\rangle_{D\bar{D}\to\phi_1\phi_1}(T_1)\left(Y^2_D-Y_D^{\rm eq}(T_1)^2\frac{Y^2_{\phi_1}}{Y^{\rm eq}_{\phi_1}(T_1)^2}\right)\nonumber \\
&+\langle\sigma v\rangle_{f\bar{f}\to\phi_1}(T)\left(Y_{q}^{\rm eq}(T)^2-Y_{q}^{\rm eq}(T)^2\frac{Y_{\phi_1}}{Y^{\rm eq}_{\phi_1}(T_1)}\right)-\langle\sigma v\rangle_{\phi_1\phi_1\to\phi_2\phi_2}(T_1)\left(Y^2_{\phi_1}-Y_{\phi_1}^{\rm eq}(T_1)^2\frac{Y^2_{\phi_2}}{Y^{\rm eq}_{\phi_2}(T_2)^2}\right)\nonumber \\
&+\langle\sigma v\rangle_{\phi_3\phi_3\to\phi_1\phi_1}(T_3)\left(Y^2_{\phi_3}-Y_{\phi_3}^{\rm eq}(T_3)^2\frac{Y^2_{\phi_1}}{Y^{\rm eq}_{\phi_1}(T_1)^2}\right)-\langle\sigma v\rangle_{\phi_1\phi_1\to\phi_2}(T_1)\left(Y_{\phi_1}^2-Y_{\phi_1}^{\rm eq}(T_1)^2\frac{Y_{\phi_2}}{Y^{\rm eq}_{\phi_2}(T_2)}\right),
\end{align}
\begin{align}
\label{yp2}
\mathcal{Q}_{\phi_2}=&\langle\sigma v\rangle_{\phi_1\phi_1\to\phi_2\phi_2}(T_1)\left(Y^2_{\phi_1}-Y_{\phi_1}^{\rm eq}(T_1)^2\frac{Y^2_{\phi_2}}{Y^{\rm eq}_{\phi_2}(T_2)^2}\right)+\langle\sigma v\rangle_{\phi_1\phi_1\to\phi_2}(T_1)\left(Y_{\phi_1}^2-Y_{\phi_1}^{\rm eq}(T_1)^2\frac{Y_{\phi_2}}{Y^{\rm eq}_{\phi_2}(T_2)}\right) \nonumber \\
&+\langle\sigma v\rangle_{\phi_3\phi_3\to\phi_2\phi_2}(T_3)\left(Y^2_{\phi_3}-Y_{\phi_3}^{\rm eq}(T_3)^2\frac{Y^2_{\phi_2}}{Y^{\rm eq}_{\phi_2}(T_2)^2}\right)+\langle\sigma v\rangle_{\phi_3\phi_3\to\phi_2}(T_3)\left(Y_{\phi_3}^2-Y_{\phi_3}^{\rm eq}(T_3)^2\frac{Y_{\phi_2}}{Y^{\rm eq}_{\phi_2}(T_2)}\right),
\end{align}
\begin{align}
\label{yp3}
\mathcal{Q}_{\phi_3}=&-\langle\sigma v\rangle_{\phi_3\phi_3\to\phi_2\phi_2}(T_3)\left(Y^2_{\phi_3}-Y_{\phi_3}^{\rm eq}(T_3)^2\frac{Y^2_{\phi_2}}{Y^{\rm eq}_{\phi_2}(T_2)^2}\right)-\langle\sigma v\rangle_{\phi_3\phi_3\to\phi_1\phi_1}(T_3)\left(Y^2_{\phi_3}-Y_{\phi_3}^{\rm eq}(T_3)^2\frac{Y^2_{\phi_1}}{Y^{\rm eq}_{\phi_1}(T_1)^2}\right)\nonumber \\
&-\langle\sigma v\rangle_{\phi_3\phi_3\to\phi_2}(T_3)\left(Y_{\phi_3}^2-Y_{\phi_3}^{\rm eq}(T_3)^2\frac{Y_{\phi_2}}{Y^{\rm eq}_{\phi_2}(T_2)}\right)
\Bigg].
\end{align}
The source terms are given by
\begin{align}
j_1&=2n_{f}^{\rm eq}(T)^2\left(1-\frac{n_D^2}{n^{\rm eq}_{D}(T_1)^2}\right)J(f\bar{f}\to D\bar{D})+2n_{q}^{\rm eq}(T)^2\left(1-\frac{n_{\phi_1}^2}{n^{\rm eq}_{\phi_1}(T_1)^2}\right)J(f\bar{f}\to \phi_1\phi_1) \nonumber \\
&+n_{q}^{\rm eq}(T)^2 J(f\bar f\to\phi_1)-n_{\phi_1}J(\phi_1\to f\bar f)+2\left(n_{\phi_2}-n_{\phi_2}^{\rm eq}\frac{n_{\phi_1}^2}{n^{\rm eq}_{\phi_1}(T_1)^2}\right)J(\phi_2\to\phi_1\phi_1) \nonumber \\
&-2\left(n_{\phi_1}^2-n_{\phi_1}^{\rm eq}(T_1)^2\frac{n_{\phi_2}^2}{n^{\rm eq}_{\phi_2}(T_2)^2}\right)J(\phi_1\phi_1\to\phi_2\phi_2) \nonumber \\
&-2\left(n_{\phi_1}^2-n_{\phi_1}^{\rm eq}(T_1)^2\frac{n_{\phi_3}^2}{n^{\rm eq}_{\phi_3}(T_3)^2}\right)J(\phi_1\phi_1\to\phi_3\phi_3),
\end{align}
\begin{align}
j_2&=-\left(n_{\phi_2}-n_{\phi_2}^{\rm eq}\frac{n_{\phi_1}^2}{n^{\rm eq}_{\phi_1}(T_1)^2}\right)J(\phi_2\to\phi_1\phi_1)+2\left(n_{\phi_1}^2-n_{\phi_1}^{\rm eq}(T_1)^2\frac{n_{\phi_2}^2}{n^{\rm eq}_{\phi_2}(T_2)^2}\right)J(\phi_1\phi_1\to\phi_2\phi_2) \nonumber \\
&+2\left(n_{\phi_3}^2-n_{\phi_3}^{\rm eq}(T_3)^2\frac{n_{\phi_2}^2}{n^{\rm eq}_{\phi_2}(T_2)^2}\right)J(\phi_3\phi_3\to\phi_2\phi_2)-\left(n_{\phi_2}-n_{\phi_2}^{\rm eq}\frac{n_{\phi_3}^2}{n^{\rm eq}_{\phi_3}(T_3)^2}\right)J(\phi_2\to\phi_3\phi_3),
\end{align}
\begin{align}
j_3&=-2\left(n_{\phi_3}^2-n_{\phi_3}^{\rm eq}(T_3)^2\frac{n_{\phi_2}^2}{n^{\rm eq}_{\phi_2}(T_2)^2}\right)J(\phi_3\phi_3\to\phi_2\phi_2)+2\left(n_{\phi_2}-n_{\phi_2}^{\rm eq}\frac{n_{\phi_3}^2}{n^{\rm eq}_{\phi_3}(T_3)^2}\right)J(\phi_2\to\phi_3\phi_3) \nonumber \\
&+2\left(n_{\phi_1}^2-n_{\phi_1}^{\rm eq}(T_1)^2\frac{n_{\phi_3}^2}{n^{\rm eq}_{\phi_3}(T_3)^2}\right)J(\phi_1\phi_1\to\phi_3\phi_3).
\end{align}
In the above we have
\begin{align}
n_f^{\rm eq}(T)^2 J(f\bar{f}\to XX)(T)&=\frac{T}{32\pi^4}\int_{4m^2_D}^{\infty}ds~\sigma_{XX\to f\bar{f}}s(s-s_0)K_2(\sqrt{s}/T), \\
n_q^{\rm eq}(T)^2 J(f\bar{f}\to \phi_1)(T)&=\frac{T}{32\pi^4}\int_{s_0}^{\infty}ds~\sigma_{f\bar{f}\to \phi_1}s(s-s_0)K_2(\sqrt{s}/T), \\
n_X(T_i)^2J(XX\to YY)(T_i)&=\frac{n_X(T_i)^2}{8m^4_X T_i K_2^2(m_X/T_i)}\int_{s_0}^{\infty}ds~\sigma_{XX\to YY}s(s-s_0)K_2(\sqrt{s}/T_i), \\
n_{X}J(X\to YY)(T_i)&=n_{X}m_{X}\Gamma_{X\to YY}\,.
\label{n2J}
\end{align}


\begin{thebibliography}{999}
\bibitem{Ackerman:2008kmp}
L.~Ackerman, M.~R.~Buckley, S.~M.~Carroll and M.~Kamionkowski,
Phys. Rev. D \textbf{79}, 023519 (2009)
doi:10.1103/PhysRevD.79.023519
[arXiv:0810.5126 [hep-ph]].

\bibitem{Heurtier:2019git}
L.~Heurtier, Y.~Mambrini and M.~Pierre,
Phys. Rev. D \textbf{99}, no.9, 095014 (2019)
doi:10.1103/PhysRevD.99.095014
[arXiv:1902.04584 [hep-ph]].

\bibitem{Aboubrahim:2020lnr}
A.~Aboubrahim, W.~Z.~Feng, P.~Nath and Z.~Y.~Wang,
Phys. Rev. D \textbf{103}, no.7, 075014 (2021)
doi:10.1103/PhysRevD.103.075014
[arXiv:2008.00529 [hep-ph]].

\bibitem{Aboubrahim:2021ycj}
A.~Aboubrahim, W.~Z.~Feng, P.~Nath and Z.~Y.~Wang,
JHEP \textbf{06}, 086 (2021)
doi:10.1007/JHEP06(2021)086
[arXiv:2103.15769 [hep-ph]].

\bibitem{Bowman:2018yin}
J.~D.~Bowman, A.~E.~E.~Rogers, R.~A.~Monsalve, T.~J.~Mozdzen and N.~Mahesh,
Nature \textbf{555}, no.7694, 67-70 (2018)
doi:10.1038/nature25792
[arXiv:1810.05912 [astro-ph.CO]].

\bibitem{Aboubrahim:2021ohe}
A.~Aboubrahim, P.~Nath and Z.~Y.~Wang,
JHEP \textbf{12}, 148 (2021)
doi:10.1007/JHEP12(2021)148
[arXiv:2108.05819 [hep-ph]].

\bibitem{Munoz:2018pzp}
J.~B.~Mu\~noz and A.~Loeb,
Nature \textbf{557}, no.7707, 684 (2018)
doi:10.1038/s41586-018-0151-x
[arXiv:1802.10094 [astro-ph.CO]].

\bibitem{Munoz:2018jwq}
J.~B.~Mu\~noz, C.~Dvorkin and A.~Loeb,
Phys. Rev. Lett. \textbf{121}, no.12, 121301 (2018)
doi:10.1103/PhysRevLett.121.121301
[arXiv:1804.01092 [astro-ph.CO]].

\bibitem{Liu:2019knx}
H.~Liu, N.~J.~Outmezguine, D.~Redigolo and T.~Volansky,
Phys. Rev. D \textbf{100}, no.12, 123011 (2019)
doi:10.1103/PhysRevD.100.123011
[arXiv:1908.06986 [hep-ph]].

\bibitem{Berlin:2018sjs}
A.~Berlin, D.~Hooper, G.~Krnjaic and S.~D.~McDermott,
Phys. Rev. Lett. \textbf{121}, no.1, 011102 (2018)
doi:10.1103/PhysRevLett.121.011102
[arXiv:1803.02804 [hep-ph]].

\bibitem{Planck:2018vyg}
N.~Aghanim \textit{et al.} [Planck],
Astron. Astrophys. \textbf{641}, A6 (2020)
[erratum: Astron. Astrophys. \textbf{652}, C4 (2021)]
doi:10.1051/0004-6361/201833910
[arXiv:1807.06209 [astro-ph.CO]].

\bibitem{Aghanim:2018eyx}
N.~Aghanim \textit{et al.} [Planck],
Astron. Astrophys. \textbf{641}, A6 (2020)
[erratum: Astron. Astrophys. \textbf{652}, C4 (2021)]
doi:10.1051/0004-6361/201833910
[arXiv:1807.06209 [astro-ph.CO]].

\bibitem{Foot:2014uba}
R.~Foot and S.~Vagnozzi,
Phys. Rev. D \textbf{91}, 023512 (2015)
doi:10.1103/PhysRevD.91.023512
[arXiv:1409.7174 [hep-ph]].

\bibitem{Foot:2016wvj}
R.~Foot and S.~Vagnozzi,
JCAP \textbf{07}, 013 (2016)
doi:10.1088/1475-7516/2016/07/013
[arXiv:1602.02467 [astro-ph.CO]].

\bibitem{Rothstein:2001tu}
I.~Rothstein and W.~Skiba,
Phys. Rev. D \textbf{65}, 065002 (2002)
doi:10.1103/PhysRevD.65.065002
[arXiv:hep-th/0109175 [hep-th]].

\bibitem{Douglas:1996sw}
M.~R.~Douglas and G.~W.~Moore,
[arXiv:hep-th/9603167 [hep-th]].

\bibitem{Arkani-Hamed:2001kyx}
N.~Arkani-Hamed, A.~G.~Cohen and H.~Georgi,
Phys. Rev. Lett. \textbf{86}, 4757-4761 (2001)
doi:10.1103/PhysRevLett.86.4757
[arXiv:hep-th/0104005 [hep-th]].

\bibitem{Hill:2000mu}
C.~T.~Hill, S.~Pokorski and J.~Wang,
Phys. Rev. D \textbf{64}, 105005 (2001)
doi:10.1103/PhysRevD.64.105005
[arXiv:hep-th/0104035 [hep-th]].

\bibitem{Aboubrahim:2021dei}
A.~Aboubrahim, W.~Z.~Feng, P.~Nath and Z.~Y.~Wang,
[arXiv:2106.06494 [hep-ph]].

\bibitem{Holdom:1985ag}
B.~Holdom,
Phys. Lett. B \textbf{166}, 196-198 (1986)
doi:10.1016/0370-2693(86)91377-8

\bibitem{Kors:2004dx}
B.~Kors and P.~Nath,
Phys. Lett. B \textbf{586}, 366-372 (2004)
doi:10.1016/j.physletb.2004.02.051
[arXiv:hep-ph/0402047 [hep-ph]].

\bibitem{Feldman:2007wj}
D.~Feldman, Z.~Liu and P.~Nath,
Phys. Rev. D \textbf{75}, 115001 (2007)
doi:10.1103/PhysRevD.75.115001
[arXiv:hep-ph/0702123 [hep-ph]].

\bibitem{Cheung:2007ut}
K.~Cheung and T.~C.~Yuan,
JHEP \textbf{03}, 120 (2007)
doi:10.1088/1126-6708/2007/03/120
[arXiv:hep-ph/0701107 [hep-ph]].

\bibitem{Aboubrahim:2019kpb}
A.~Aboubrahim, W.~Z.~Feng and P.~Nath,
JHEP \textbf{02}, 118 (2020)
doi:10.1007/JHEP02(2020)118
[arXiv:1910.14092 [hep-ph]].

\bibitem{Du:2022fqv}
M.~Du, Z.~Liu and P.~Nath,
[arXiv:2204.09024 [hep-ph]].

\bibitem{Aboubrahim:2020afx}
A.~Aboubrahim, T.~Ibrahim, M.~Klasen and P.~Nath,
Eur. Phys. J. C \textbf{81}, no.8, 680 (2021)
doi:10.1140/epjc/s10052-021-09483-0
[arXiv:2012.10795 [hep-ph]].

\bibitem{PandaX-II:2021lap}
J.~Yang \textit{et al.} [PandaX-II],
Sci. China Phys. Mech. Astron. \textbf{64}, no.11, 11 (2021)
doi:10.1007/s11433-021-1740-2
[arXiv:2104.14724 [hep-ex]].

\bibitem{Huo:2017vef}
R.~Huo, M.~Kaplinghat, Z.~Pan and H.~B.~Yu,
Phys. Lett. B \textbf{783}, 76-81 (2018)
doi:10.1016/j.physletb.2018.06.024
[arXiv:1709.09717 [hep-ph]].

\bibitem{Hall:2009bx}
L.~J.~Hall, K.~Jedamzik, J.~March-Russell and S.~M.~West,
JHEP \textbf{03}, 080 (2010)
doi:10.1007/JHEP03(2010)080
[arXiv:0911.1120 [hep-ph]].

\bibitem{Pitrou:2018cgg}
C.~Pitrou, A.~Coc, J.~P.~Uzan and E.~Vangioni,
Phys. Rept. \textbf{754}, 1-66 (2018)
doi:10.1016/j.physrep.2018.04.005
[arXiv:1801.08023 [astro-ph.CO]].

\bibitem{Planck:2018nkj}
N.~Aghanim \textit{et al.} [Planck],
Astron. Astrophys. \textbf{641}, A1 (2020)
doi:10.1051/0004-6361/201833880
[arXiv:1807.06205 [astro-ph.CO]].

\bibitem{Riess:2021jrx}
A.~G.~Riess, W.~Yuan, L.~M.~Macri, D.~Scolnic, D.~Brout, S.~Casertano, D.~O.~Jones, Y.~Murakami, L.~Breuval and T.~G.~Brink, \textit{et al.}
[arXiv:2112.04510 [astro-ph.CO]].

\bibitem{Riess:2018uxu}
A.~G.~Riess, S.~Casertano, W.~Yuan, L.~Macri, J.~Anderson, J.~W.~MacKenty, J.~Bradley Bowers, K.~I.~Clubb, A.~V.~Filippenko and D.~O.~Jones, \textit{et al.}
Astrophys. J. \textbf{855}, no.2, 136 (2018)
doi:10.3847/1538-4357/aaadb7
[arXiv:1801.01120 [astro-ph.SR]].

\bibitem{Riess:2019cxk}
A.~G.~Riess, S.~Casertano, W.~Yuan, L.~M.~Macri and D.~Scolnic,
Astrophys. J. \textbf{876}, no.1, 85 (2019)
doi:10.3847/1538-4357/ab1422
[arXiv:1903.07603 [astro-ph.CO]].

\bibitem{Abdalla:2022yfr}
E.~Abdalla, G.~Franco Abell\'an, A.~Aboubrahim, A.~Agnello, O.~Akarsu, Y.~Akrami, G.~Alestas, D.~Aloni, L.~Amendola and L.~A.~Anchordoqui, \textit{et al.}
JHEAp \textbf{34}, 49-211 (2022)
doi:10.1016/j.jheap.2022.04.002
[arXiv:2203.06142 [astro-ph.CO]].

\bibitem{Aboubrahim:2022gjb}
A.~Aboubrahim, M.~Klasen and P.~Nath,
JCAP \textbf{04}, no.04, 042 (2022)
doi:10.1088/1475-7516/2022/04/042
[arXiv:2202.04453 [astro-ph.CO]].

\bibitem{CMB-S4:2016ple}
K.~N.~Abazajian \textit{et al.} [CMB-S4],
[arXiv:1610.02743 [astro-ph.CO]].

\bibitem{Abazajian:2022nyh}
K.~Abazajian \textit{et al.} [CMB-S4],
[arXiv:2203.08024 [astro-ph.CO]].

\bibitem{Hindmarsh:2005ix}
M.~Hindmarsh and O.~Philipsen,
Phys. Rev. D \textbf{71}, 087302 (2005)
doi:10.1103/PhysRevD.71.087302
[arXiv:hep-ph/0501232 [hep-ph]].
\end{thebibliography}
\end{document}